\documentclass[journal,comsoc]{IEEEtran}
\usepackage{algorithm}
\usepackage{algpseudocode}
\usepackage[utf8]{inputenc}
\usepackage{graphicx}
\usepackage{cite}
\graphicspath{{figs/}}
\usepackage{float}
\usepackage{amsmath}

\usepackage[normalem]{ulem}
\useunder{\uline}{\ul}{}
\usepackage{multirow}
\usepackage{lipsum}
\usepackage{textcomp}
\usepackage{comment}
\usepackage{xcolor}
\usepackage{gensymb}
\usepackage{tikz}
\usepackage{tikz-3dplot, filecontents}
\usetikzlibrary{shapes.geometric, arrows, backgrounds,math, calc}
\usetikzlibrary{automata,positioning}
\usetikzlibrary{shapes.misc}
\usetikzlibrary{circuits.ee.IEC}
\usepackage{circuitikz}
\usepackage{pgfplots}
\usepackage{amssymb}
\pgfplotsset{compat=1.17}
\usepgfplotslibrary{colorbrewer}  % for color
\pgfplotsset{minor grid style={dashed}}
\title{Towards a Sustainable Internet-of-Underwater-Things based on AUVs, SWIPT and Reinforcement Learning\\

\thanks{This work was supported by the Petroleum Technology Development Fund (PTDF) of the Federal Republic of Nigeria [grant number 1353/18].}
\thanks{The authors are with the James Watt School of Engineering, University of Glasgow, U.K. (E-mail: \{Kenechi.Omeke, Michael.Mollel, SyedTariq.Shah, Lei.Zhang, Qammer.Abbasi, Muhammad.Imran\}@glasgow.ac.uk). \\
Corresponding author: Kenechi Omeke (Kenechi.Omeke@glasgow.ac.uk).}
}
\author{Kenechi G. Omeke,~\IEEEmembership{Graduate Student Member,~IEEE,} Michael Mollel, Syed T. Shah, Lei Zhang,~\IEEEmembership{Senior Member,~IEEE,} Qammer H. Abbasi,~\IEEEmembership{Senior Member,~IEEE,} and Muhammad Ali Imran,~\IEEEmembership{Fellow,~IEEE}
}
\begin{document}

\maketitle
\begin{abstract}
Life on earth depends on healthy oceans, which supply a large percentage of the planet's oxygen, food, and energy.
However, the oceans are under threat from climate change, which is devastating the marine ecosystem and the economic and social systems that depend on it.
The Internet-of-underwater-things~(IoUTs), a global interconnection of underwater objects, enables round-the-clock monitoring of the oceans.
It provides high-resolution data for training machine learning~(ML) algorithms for rapidly evaluating potential climate change solutions and speeding up decision-making.
The sensors in conventional IoUTs are battery-powered, which limits their lifetime, and constitutes environmental hazards when they die.
In this paper, we propose a sustainable scheme to improve the throughput and lifetime of underwater networks, enabling them to potentially operate indefinitely.
The scheme is based on simultaneous wireless information and power transfer~(SWIPT) from an autonomous underwater vehicle~(AUV) used for data collection.
We model the problem of jointly maximising throughput and harvested power as a Markov Decision Process~(MDP), and develop a model-free reinforcement learning~(RL) algorithm as a solution.
The model's reward function incentivises the AUV to find optimal trajectories that maximise throughput and power transfer to the underwater nodes while minimising energy consumption.
To the best of our knowledge, this is the first attempt at using RL to ensure sustainable underwater networks via SWIPT.
The scheme is implemented in an open 3D RL environment specifically developed in MATLAB for this study.
The performance results show up 207\% improvement in energy efficiency compared to those of a random trajectory scheme used as a baseline model.
\end{abstract}
% , while updating its decisions according to real-time changes in the underwater environment.


\maketitle

\section{Introduction} \label{Intro}
The earth is a water planet. Over 70\% of the earth's surface is covered by water, which provides the planet with food, and energy, and regulates global temperatures and wind.
Most importantly, the oceans generate about 50\% of the oxygen used on earth and absorb about 25\% of all atmospheric carbons.
% The oceans also supply a large percentage of the energy used on land and play an increasing role in the energy mix to meet rising energy demands to power growing populations and economies.
Man's very existence and survival on earth depend on healthy oceans.
However, the oceans are under threat from pollution and climate change, which are devastating them and the economic and social systems that depend on them, leading to extreme and unpredictable weather events such as hurricanes, wildfires, flooding, and droughts around the globe.
% There is an increasing effort to understand the mechanisms of marine climate change so as to mitigate them.
% However, the oceans are vast and vary from place to place, necessitating solutions that have both a global outlook and local efficacy.
% Artificial intelligence (AI) and m
Machine learning (ML) provides tools for rapidly searching and testing potential climate solutions, but they require vast amounts of data to train their algorithms.
% : to predict and analyse the impacts of climate change, model the marine climate or identify patterns in large datasets related to climate and weather.
Conventional marine research tools that use ocean-going vessels and in-situ data analysis or remote sensing have limitations such as large delays, limited coverage, poor spatial resolution, etc.
% in the sensing area, inability to handle large datasets or rapidly changing times series data, etc.

Wireless underwater sensor networks (WUSNs) provide round-the-clock data collection at higher spatial and temporal resolutions than is possible via any other means of underwater data collection.
They are the foundation of the internet-of-underwater-things (IoUTs), whereby sensors and underwater objects are networked to cover as much of the oceans as possible~\cite{OverviewIoUTsDomingo2012}.
They provide vast amounts of data for training ML models for environmental and climate change research, industrial applications, and proactive early disaster prevention/early warning systems.
The ML models are used to automatically learn patterns in the data collected by underwater networks to improve decision-making~\cite{IoUTsBigMarineDataJahanbakht2021} and to automate underwater network operations, and enhance their performance and energy efficiency~\cite{RLIoUTs}.
% collected from underwater networks for these applications, but also
% Though AI solutions are data-hungry, they can be combined with WSNs to enable real-time monitoring of the oceans and rapid analysis of marine data~\cite{MLApplicationOceansDataLou2021} to speed up potential climate solutions and scale up the number and variety of efforts.
% Such cognitive networking underpinned by ML is revolutionising the internet-of-underwater-things (IoUTs), whereby sensors and underwater objects are networked to cover as much of the oceans as possible~\cite{OverviewIoUTsDomingo2012}, with adaptive networking decisions made in near real-time to enhance network performance and improve energy efficiency~\cite{RLIoUTs}.
% to provide faster response to marine climate emergencies as well as predict future events of interest.
However, WUSNs nodes are energy-constrained, and their operational life is limited by the size of their onboard batteries.
In addition, the batteries become environmental hazards when they die, thereby exacerbating the problem they were designed to solve.
% While there is a dire need to plant as many sensors in the oceans to monitor physico-biochemical processes therein, chemical batteries are a shortsighted solution since they can neither be replaced, recharged nor recovered when they die due to their inaccessibility and the heavy costs involved.

Energy efficiency is the most important factor limiting WUSNs due to the high energy required for data transmission and the difficulty of replacing depleted sensor batteries.
In light of the foregoing, there is an urgent need for sustainable solutions to save the oceans.
% which do not rely on the use of chemical-based batteries to power underwater sensor nodes.
% \hl{Supercapacitors are a viable solution}
Underwater energy harvesting and simultaneous wireless and information transfer (SWIPT)~\cite{SWIPTAnalysisBereketli2012} remove the need for large batteries in underwater sensor nodes and enable them to potentially operate indefinitely.
Acoustic-based SWIPT enables contact-less recharging of underwater sensor nodes at much longer distances than is possible through RF, optical communication, or magnetic induction~\cite{BatterylessIoUTGuida2022}.
% Underwater simultaneous wireless information and power transfer (SWIPT)~\cite{SWIPTAnalysisBereketli2012} is gaining favourable attention IoUT networking as one such solution.
% Underwater sensor networks rely on acoustic telemetry since RF, optical and magnetic induction solutions have severely limited transmission ranges.
In underwater SWIPT, the signal transmitted by an acoustic source is used to decode data at the receivers and to charge a bank of supercapacitors for powering WUSNs nodes~\cite{BatterylessIoUTGuida2022, SWIPTAnalysisBereketli2012}.
Supercapacitors are lighter than batteries, charge faster, and pose fewer environmental problems~\cite{BatterylessIoUTGuida2022}.

% \begin{figure*}[htbp]
% \centering
% \includegraphics[width=2.5\columnwidth]{figs/auv_parts.png}
% \caption{AUV parts for inspiration.}
% \label{fig:auv-model}
% \end{figure*}

% \begin{figure}[htbp]
% \centering
% \includegraphics[width=\columnwidth]{figs/auv_img.png}
% \caption{Image to be changed -- AUV parts}
% \label{fig:auv-model}
% \end{figure}

This paper proposes a sustainable underwater sensor network based on SWIPT and AUVs. 
The AUVs are deployed from a floating platform on the water surface to simultaneously collect data from the underwater nodes and recharge the nodes.
% The AUV is deployed from a floating platform on the water surface to a given depth, from where it serves as a gateway for the network of underwater sensor nodes below it.
Each AUV is equipped with a bidirectional acoustic modem; one for communicating with the underwater sensor nodes and the other for communicating with the floating station on the water surface.
It is imperative to improve the throughput of the system to maximise data collection and at the same time, maximise WPT to the nodes.
% the magnitude of harvested.
However, finding locations that simultaneously maximise these goals is a non-trivial problem due to the three-dimensional nature of the ocean, and water current, which causes the AUVs to sway and nodes to drift.
In acoustic-based SWIPT, an external acoustic source generates data-bearing acoustic waves that can be picked up by a piezoelectric ceramic transducer some distance away from the source, which
% propagate away from the source. At a receiver  a piezoelectric ceramic transducer can be used to recover both the data encoded in the arriving acoustic waves and 
extracts electrical signal from the waves, for data decoding and to charge the transducer's power source (a bank of supercapacitors).
Acoustic SWIPT 
% can indefinitely extend the lifetime of underwater networks, the technology 
is still in its infancy due to the severe limitations such as heavy channel losses due to absorption and spreading~\cite{CapacityDistanceStojanovic2007},
% which implies that only a fraction of the radiated acoustic power reaches the receiver, 
as well as high electrical--acoustic and acoustic--electrical conversion losses at the transmitter and receiver~\cite{BatterylessIoUTGuida2022}, respectively.

There is a dearth of current literature in acoustic-based SWIPT, despite the existence of analytical proof of its feasibility in underwater sensor networks~\cite{SWIPTAnalysisBereketli2012, AcousticContactlessChargingShahab2015} and its practical demonstration, as shown in~\cite{BatterylessIoUTGuida2022, WirelessAcousticTransferKim2022, AcousticContactlessChargingShahab2015, PracticalAcousticWPTSi2013}.
The pioneering work in acoustic SWIPT was presented in~\cite{SWIPTAnalysisBereketli2012}, which showed analytical proof of the feasibility of WPT to WUSNs nodes from a given distance.
The authors derived the theoretical upper bounds for harvested power
% electrical power for an array of receiver hydrophones 
for a given underwater channel, source power budget, known source and receiver characteristics, and range. 
% (receiver sensitivity, electrical to acoustic conversion efficiency and vice versa) 
% They also investigated how the directivity of the source affects the magnitude of harvested power and network coverage.
They showed that up to 100 W of electrical power can be harvested for an input electrical power of 2 kW for frequencies less than 20 kHz at less than 1 km distances.
A batteryless underwater system was proposed in~\cite{BatterylessIoUTGuida2022}, where it was demonstrated that sufficient power can be harvested via SWIPT to operate an underwater sensor node for both sensing and communication operations, without an external power supply.
The authors in~\cite{AcousticContactlessChargingShahab2015, UltrasoundEnergyTransferShahab2014} quantitatively and experimentally evaluated the magnitude of electrical power delivered to a remote node from a source of known acoustic strength, as well as the parameters that influence the received power, such as the transmission range, source strength, and the impedance characteristics of the receiver.
They used a receiver comprising a piezoelectric cylindrical bar operating in the 33-mode (longitudinally excited) of piezoelectricity under free-free mechanical boundary conditions, while a spherical wave generator was used as a source.

Other sustainable solutions proposed for sustainable WUSNs include energy harvesting via scavenging energy from the ambient environment, which can power only a single node at a time, and WPT through inductive and capacitive coupling.
Compared to energy scavenging, an underwater acoustic source can simultaneously power multiple underwater nodes via SWIPT~\cite{SWIPTAnalysisBereketli2012}.
The authors in~\cite{BatterylessIoUTGuida2022} also showed that hardware reuse is another advantage of acoustic-based underwater SWIPT over energy scavenging, as the same transducer hardware used for communication can be reused for recharging the energy sources of nodes.
%Energy harvesting for time-varying systems has been modelled as an MDP in~\cite{MDPsEHFadingChannelsLi2015}; however, the authors assumed that the transition probabilities of the system are known, which is unrealistic in underwater networks.
% In~\cite{RLTidalHarvestingHan2020}, it was shown that the arrival of energy harvesting signal and traffic demands by nodes in an underwater network is stochastic due to the diverse long propagation delays of multipath acoustic waves and location uncertainty induced by node mobility arising from water motion.
% This necessitates medium access solutions that are dynamic and which adapt to changes in the network in an online manner.
% RL is well suited to solving such dynamic problems, especially where the global network information is unavailable to the network nodes.
% The authors in~\cite{RLTidalHarvestingHan2020} investigated the impact of the slow speed of acoustic waves and water motion on underwater network performance in terms of throughput and fairness in accessing the network resources.
% They formulated the problem of selecting an optimal window size for channel access for the network nodes as a multi-armed bandit problem, where different contention window sizes are represented as arms of the bandit. By interacting with the network environment, the learning agent (software at each node) learns the optimal window size to use based on ACK feedback received from the access point used as a base station.
In~\cite{RLTidalHarvestingHan2020}, multi-armed bandit RL was proposed to aid tidal energy harvesting for an IoUTs network, 
% whereby channel access window sizes for channel access for the network nodes as a multi-armed bandit problem, 
where different channel access contention window sizes were represented as arms of the bandit. The RL agent learns the optimal window size through interaction with the network.
A shortest path charging scheme based on $k-$Means clustering was proposed in~\cite{ShortestPathChargingAUVLin2018} to improve throughput and minimise the travel distance for mobile robots used to recharge underwater nodes' batteries, but it lacked a channel model for evaluating the signalling technology used. 
% However, the authors did not discuss the underlying channel model nor indicate the communication technology used.
An acoustic modem was designed and tested in~\cite{UltrasonicTransducerDesignZhao2021} for SWIPT, which considers how to achieve high transmission efficiency underwater. 
Power transfer was maximised through electrical impedance matching (using a resonance compensation circuit) and acoustic impedance matching (by covering the bakelite shell with an iron shell to reduce energy leakage).
% To achieve electrical impedance matching, the authors added a resonance compensation circuit; to achieve acoustic impedance matching, they modified the transducer structure (by covering the bakelite shell with an iron shell) to reduce radial vibrations, which causes radial energy leakage.
The work in~\cite{AcousticSWIPTNOMAEsmaiel2020} considered acoustic-based SWIPT for unmanned underwater vehicles.
% However, the authors assumed a linear energy harvesting model in which the entire acoustic energy incident at the receiver is harvested.
However, the authors assumed a linear energy harvesting model, whereas practical SWIPT circuits are non-linear~\cite{SWIPTAnalysisBereketli2012, BatterylessIoUTGuida2022}, as we show in Section \ref{sec:swipt-physics}.

The bulk of the available literature on WPT in underwater networks is based on inductive and capacitive coupling~\cite{WPTAUVsWang2023, ReviewWirelessAUVChargingTeeneti2021, ConductiveCouplerUnderwaterTamura2021,  MIMultiAUVChargingGuo2021, InductiveCouplingWPTLin2019, SEANetDemirors2016, InductivePowerTransferCheng2014 }. Inductive coupling is based on electromagnetic induction, whereby two conductors are configured to induce a voltage in one conductor due to a change in electric current in the other.
In underwater applications, WPT based on inductive coupling is achieved using spiral or coil elements.
These techniques can generate milli-watts of power at a few centimetres to meters, with typical WPT efficiencies between 50\% and 80\% at centimetre distances~\cite{BatterylessIoUTGuida2022}.

In this paper, we model the problem of simultaneously maximising throughput and harvested power as a Markov Decision Process (MDP).
To deal with the stochastic and time-varying nature of the underwater acoustic channels, we propose a model-free RL scheme based on $Q-$learning and state-action-reward-state-action (SARSA) to solve the MDP.
The aim of this work is to simultaneously maximise throughput and harvested power in the underwater network by finding optimal trajectories for the AUV to follow during data collection rounds.
% that guarantees simultaneous coverage for multiple underwater sensor nodes involved in SWIPT.
% In terrestrial networks, this problem has received significant attention as the drone coverage/placement problem (e.g.~\cite{OptimalUAVBSTrajectoriesSaxena2019}). 
The proposed scheme considers SWIPT based on the power splitting (PS) technique, whereby a portion of the received power ($\alpha$) is used for communication and the remainder ($1-\alpha$) for WPT.
To evaluate the performance of the considered scheme, we developed a three-dimensional (3D) underwater environment in MATLAB to enable the deployment of RL algorithms. 
The environment can be used to test any model-free RL algorithm (as well as model-based schemes if a model of the environment is provided).
%In the interest of open research, we freely share the source code and the environment with the research\footnote{Due to funding constraints, access to the source code will be restricted pending publication of this article.}. 
The proposed approach is novel, and to the best of the authors' knowledge, is the first attempt at using RL to jointly optimise throughput and harvested power in a SWIPT-enabled underwater acoustic network.
The main contributions of this article are summarised as follows:
\begin{itemize}
    \item We model the joint optimisation of throughput and harvested power in a SWIPT-based underwater acoustic sensor network as an MDP and propose a solution based on model-free RL algorithms using $Q-$learning and SARSA.
    
    % We formulate a reward function to guide an RL agent towards a solution and test the performance of different model-free RL algorithms.
    \item RL  is used to find the optimal AUV trajectory that simultaneously maximises throughput and harvested power, as well as to find the optimal PS ratio, $\alpha$, for dynamically allocating the received power for data transmission and WPT.
    % \item Through interactions with the underwater environment, we find the optimal placement of the AUV that optimises coverage, throughput and harvested power.
    \item We developed an open 3D RL environment for testing RL algorithms for underwater sensor networks. The environment is independent of the underlying channel model and can be used with underwater signalling technology (acoustic, optical, magnetic induction, or radio frequency).
\end{itemize}
The rest of this paper is structured as follows.
The network model, underlying underwater channel model and physics of acoustic-based SWIPT are covered in Section \ref{sec:system-model}.
Section \ref{sec:rl-model} presents some background on RL, provides details on modelling the joint maximisation of throughput and harvested power as an MDP, and presents the RL solution.
This is followed by an evaluation of the performance of the proposed scheme and a presentation of the achieved results in Section \ref{sec:simulations}.
Section \ref{conclusion} concludes the paper and highlights future directions.

\section{System Model}\label{sec:system-model}
\subsection{Network Model}
\begin{figure}[t]
\centering
\includegraphics[width=0.98\columnwidth]{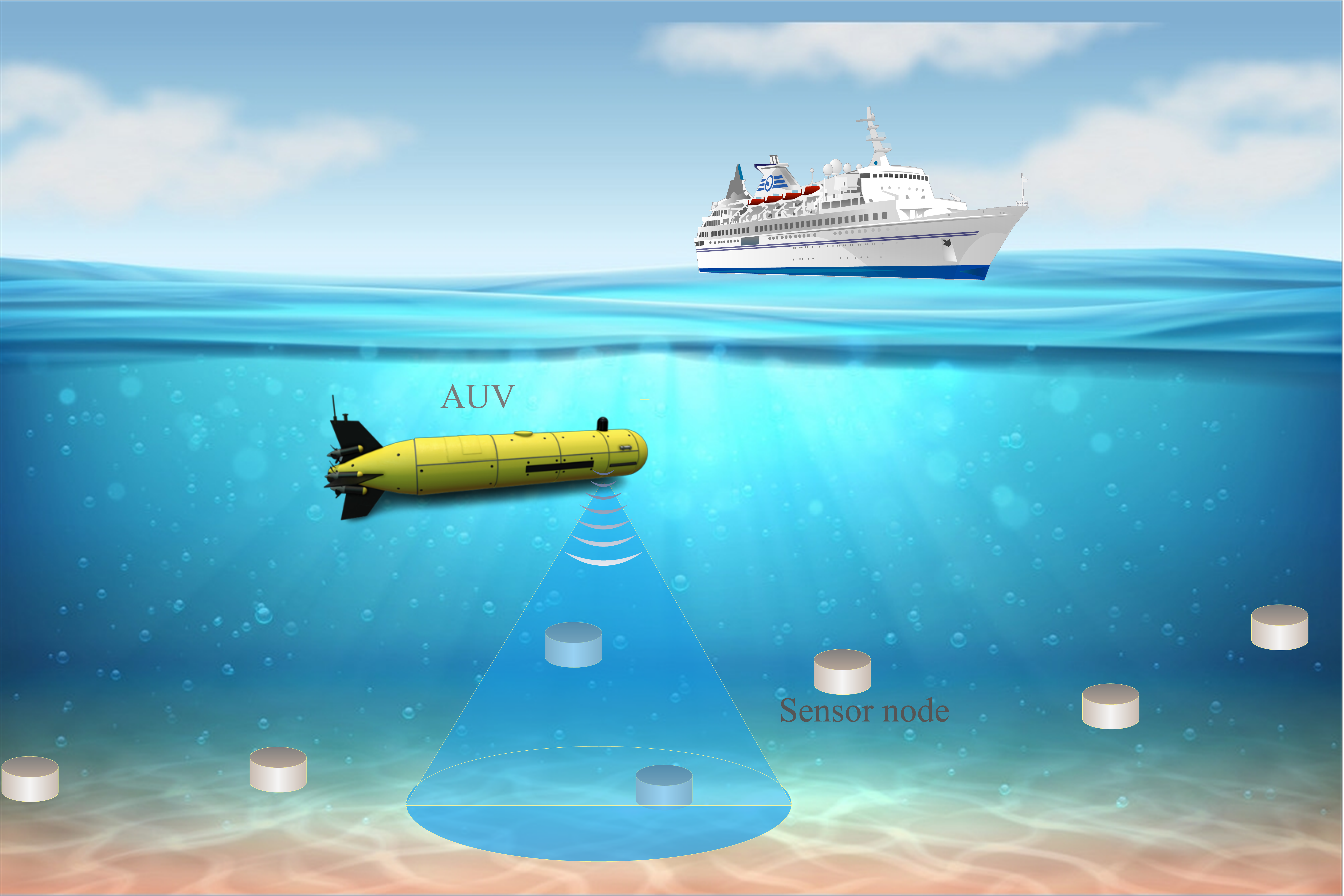}
\caption{AUV moves around the network and provides SWIPT-based energy harvesting to power-constrained underwater sensor nodes. The harvested power is used to recharge an onboard battery for sensing, data processing and communication of aggregated data from cluster heads to the AUV, which serves as a local base station.}
\label{fig:model}
\end{figure}

\begin{figure}[ht]
    \centering
    \includegraphics[width=8cm]{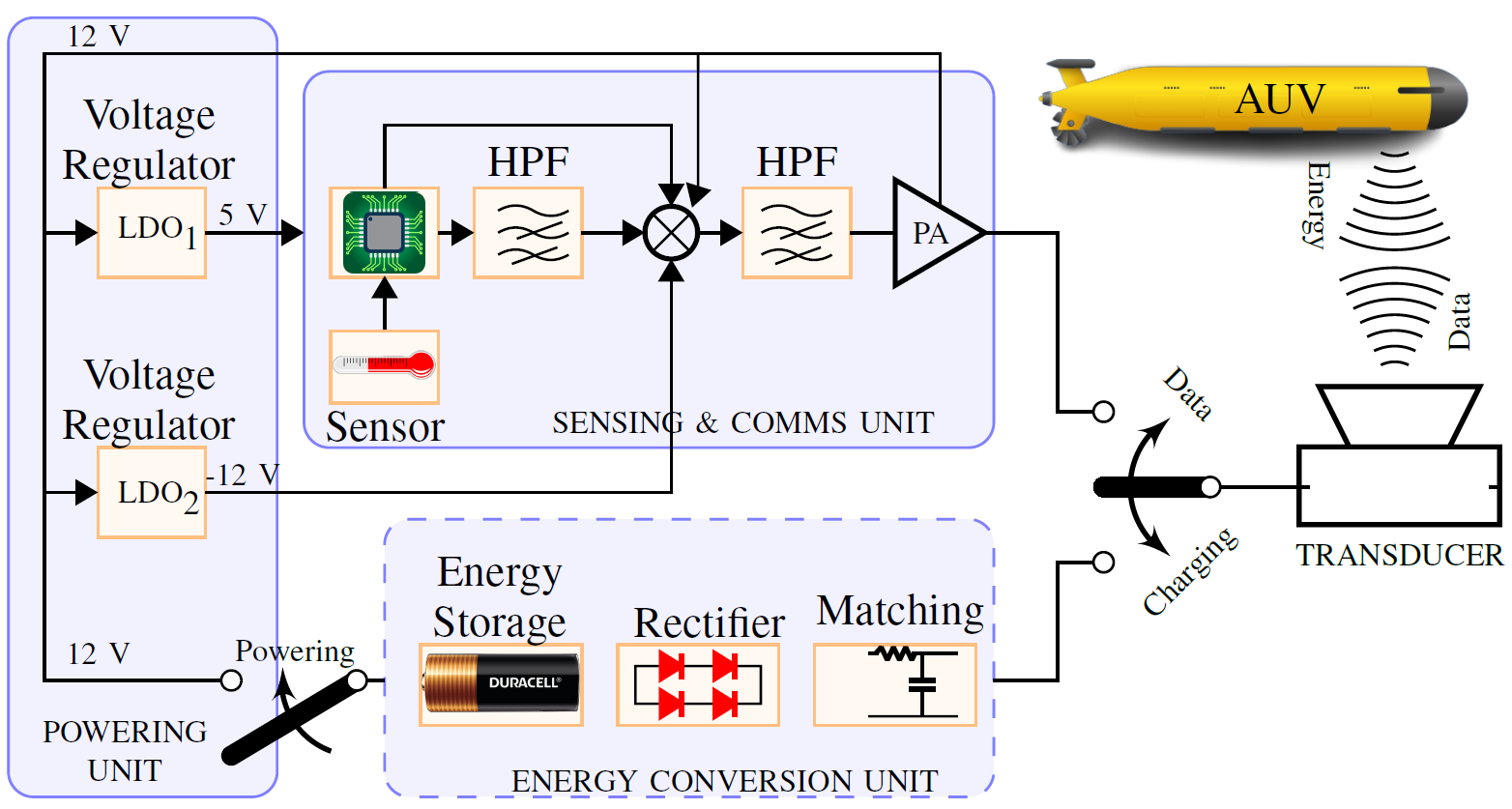}
    \caption{Block diagram of a SWIPT-powered underwater sensor system, with circuitry for electrical to acoustic power conversion and vice versa.}
    \label{fig:example}
\end{figure}

A 3D underwater sensor network is considered, as shown in Fig. \ref{fig:model}.
The environment comprises an AUV deployed from a floating station and $N$ sensor nodes distributed within a 3D volume of dimensions $L \times W \times H$, where $L$ and $W$ represent the length and width of the network, respectively, and $H$ represents the depth below sea level.
Each sensor node contains sensors for monitoring key underwater parameters such as temperature, power of hydrogen (pH) and dissolved oxygen.
% Each one is equipped with an energy management unit, a transceiver and a small microcontroller which serves as the brain of the node and performs functions related to data processing, energy management, etc.
% ~\hl{update based on actual controllers, such as SEANet}.
Each sensor node consists of a suite of sensors and devices divided into three major units for communication and sensing, energy conversion, and power management.
These units interface with the AUV for data and power transfer through the acoustic transducer, as shown in the block diagram in Fig. \ref{fig:example}.
The communication and sensing unit contains the sensors used for environmental monitoring, a microcontroller responsible for data processing, high pass filters (HPF) for signal conditioning, and transceiver operations.
The powering unit is responsible for supplying the correct voltage logic levels required by the different components of the sensor node, while the energy conversion unit is responsible for impedance matching, power rectification, and storage.
% The suite of suite of components within a sensor node interface with the AUV for data and power transfer thro
% \hl{description of the blocks in the block diagram}
% \begin{figure}[th]
% 	\centering
% 	{\resizebox{0.99\columnwidth}{!}{\input{figs/swipt_circuit.tikz}}}
% 	\caption{Block diagram of a SWIPT-powered underwater sensor node, with additional circuitry for electrical to acoustic power conversion and vice versa.}
% 	\label{fig:example}
% \end{figure}
% \vspace{-0.75em}
% An AUV deployed from a floating platform (\hl{this may have been captured in the intro}) is used for data collection and to recharge the power supply of the sensor nodes.
The AUV can communicate with the underwater sensor nodes as well as the floating station at the water surface.
The AUV communicates with the underwater nodes via an acoustic modem/hydrophone, whose coverage takes the form of a cone~\cite{SWIPTAnalysisBereketli2012}, as shown in Fig.~\ref{fig:model}.
This coverage pattern enables it to potentially recharge multiple underwater nodes simultaneously via WPT.
% An example of an electrical-acoustic converter is the Airmar P58\footnote{The data sheet can be found at https://www.airmar.com/uploads/techdata/50-200a.pdf}.

% An AUV is deployed from a boat or ship to hover over the network location to recharge the underwater sensor nodes and collect sensed data from the nodes.
% It covers a spherical region at a depth of $z$ below the water surface, providing a conical region of wireless power and communication coverage for underwater nodes.

% \hl{Whilst it is possible to have data communication while the underwater node is charging, this comes at the cost of additional complexity (plus size and weight), cost and storage as at least two transducers are required per sensor node unit, as well as an additional storage. However, this also enables an additional function (especially having an extra storage), such as sensing and data processing in the absence of the AUV. A switching circuit can be added to alternate charging between the power storages during each AUV visit.}

% \begin{figure}[th]
% 	% \centering
% 	{\resizebox{0.99\columnwidth}{!}{\input{figs/submarine2.tikz}}}
% 	\caption{AUV moves around the network and provides SWIPT-based energy harvesting to power-constrained underwater sensor nodes. The harvested power is used to recharge an onboard battery for sensing, data processing and communication of aggregated data from cluster heads to the AUV, which serves as a local base station.}
% 	\label{fig:model}
% \end{figure}

% \begin{figure}[htbp]
% \centering
% \includegraphics[width=0.98\columnwidth]{img/cone_coverage.png}
% \caption{Coverage of directional acoustic sources~\cite{SWIPTAnalysisBereketli2012}}
% \label{fig:coverage}
% \end{figure}

% \begin{figure}[th]
% 	\centering
% 	{\resizebox{0.9\columnwidth}{!}{\input{figs/coverage_nodes.tikz}}}
% 	\caption{action space}
% 	\label{fig:action_space}
% \end{figure}

\subsection{Channel Model}
An underwater acoustic transmitter is called a projector, while an array of hydrophones is typically used as receivers.
The acoustic source level of the projector expresses the amount of power radiated, analogous to the transmitting power in terrestrial network nodes.
% Typically, an array of hydrophones is used as receivers of acoustic energy radiated from a projector.
The passive SONAR equation expresses the received acoustic intensity at a distance away from the projector as a function of acoustic losses and modem characteristics.
In typical (terrestrial network) parameters, this intensity is equivalent to the received power, which can be used to calculate the received signal-to-noise ratio ($\gamma$) as~\cite{SoundPrinciplesUrick1975}
\begin{equation}
\gamma=\mathrm{SL}-\mathrm{TL}-
\mathrm{NL} + \mathrm{DI},
\label{eq:sonar}
\end{equation} 
where SL is the acoustic source level, TL characterises the transmission losses, NL is the noise level, and DI is the directivity index, which expresses the ability of the acoustic projector to focus the radiated energy in a desired direction.
% The signal-to-noise ratio ($\gamma$) depends on the frequency of the acoustic signal, the transmitting power and transmitter-receiver separation.
% For a narrowband acoustic signal (represented by a single tone of frequency $\Delta f$), this can be expressed as~\cite{CapacityDistanceStojanovic2007}
% \begin{equation} 
% \gamma(d, f)=\frac{P_{t} / A(d, f)}{N_{0}(f) \Delta f},
% \label{eq:snr}
% \end{equation}
% where $P_{t}$ is the power of the acoustic projector, $N_{0}$ is the noise power, which is a function of frequency. $A$ is the acoustic pathloss, which is given by
% \begin{equation}
% A(d, f)=A_{0} d^{k} \alpha(f)^{d},
% \end{equation}
At the projector, an electrical input signal is used to generate an acoustic signal, which is propagated as pressure waves.
The source level depends heavily on the electrical-acoustic conversion efficiency of the projector and typically varies between 20$\%$ and 70\% for practical modems~\cite{SWIPTAnalysisBereketli2012}, implying that in some cases, less than half of the input electrical power is converted into acoustic power that can be radiated outwards.
Given similar losses at the receiver and channel losses, only a fraction of the input electrical power is available for SWIPT.
The acoustic source level at the projector is given by~\cite{SWIPTAnalysisBereketli2012}
% The source level is the acoustic power radiated by the transducer by converting electrical input energy into acoustic power.
% It should be noted that there are conversion losses encountered in the course of transforming the electrical energy to the mechanical energy that bear the acoustic vibrations.
% The source level is given by
\begin{equation}
    SL=170.8+10 \log_{10}P_{elec}+10\log_{10}\eta+DI,
\end{equation}
where $P_{elec}$ is the electrical input power at the source and $\eta$ is the electrical to acoustic power conversion efficiency.
% The channel transmission loss is a function of frequency and the separation between the acoustic source and the receiver.
% It comprises an absorption loss, $\alpha$ and a spreading loss.
% Absorption losses arise due to chemical interactions between the acoustic waves and water molecules, often leading to some of the propagating wireless energy being converted to heat. 
% It depends on the frequency of the signal; higher frequencies are absorbed more rapidly than lower frequencies.
% Spreading losses arise due to the decreasing intensity of the propagating waves away from the source. As the wavefront increases, the intensity of the signal energy per unit area decreases.
The transmission loss is given by~\cite{CapacityDistanceStojanovic2007}
\begin{equation}
10 \times \text{log}TL(r, f)= k \times 10 \text{log} r + r \times 10 \text{log} \alpha(f),
\label{eq:tl}
\end{equation}
where $r$ is the distance between the transmitter and the receiver.
The absorption loss was obtained empirically by W.H. Thorp as~\cite{ThorpFormula1967}
\begin{equation}
\alpha(f)=0.11{\frac{f^2}{1+ f^{2}}} +44{\frac{f^2}{4100+f^{2}}}+2.75\cdot 10^{-4}f^{2}+0.003,  
\end{equation}
where $f$ is the operating frequency in kHz, and $k$ is referred to as the channel spreading factor (similar to the path loss exponent in terrestrial radio-based communications);
$k$ takes values of (1, 2), where $k=1$ is referred to as cylindrical spreading and $k = 2$ is referred to as spherical spreading. 
% Cylindrical spreading dominates at shallow water depths where the depth of the acoustic modem is less than the horizontal communication range, whereas spherical spreading dominates in deep waters.

The noise level represents the cumulative effect of noise in an underwater communication system.
Noise in underwater acoustic communication systems is a function of the frequency of the propagating acoustic waves. 
It comprises noise from shipping ($N_s$), water waves ($N_w$),  water turbulence ($N_{t}$), and thermal noise ($N_th$), whose power spectral densities can be expressed in dB re 1 $\mu$Pa at 1m per Hz~\cite{CapacityDistanceStojanovic2007} as
\begin{align}
N_{t}(f)=&17 - 30\log (f) \label{eq:turbulence},\\ 
N_{s}(f)=&30 \!+\! 20s \!+\! 26\log (f) \!-\! 60\log (f\!+\!0.03),\qquad\\ 
N_{w}(f)=&50 + 7.5\sqrt {w} + 20\log (f) - 40\log (f+0.4),\\ 
N_{th}(f)=&-15 + 20\log (f) \label{eq:thermal},\end{align}
where $w$ is the wind speed in m/s, $s$ is the shipping activity factor (0 for low activity and 1 for high activity) and $f$ is the frequency in kHz.
% As equations (\ref{eq:turbulence} -- \ref{eq:thermal}) show, these noise sources are frequency-dependent. 
Shipping activity noise and noise due to ocean turbulence dominate at very low and low frequencies; thermal noise is predominant at frequencies above 100 kHz, while noise due to surface waves is strongest between 100 Hz -- 100 kHz~\cite{CapacityDistanceStojanovic2007}.
% The statistical fading model of the underwater channel is modelled as a Rician distribution~\cite{RicianChannelRadosevic2009}, as this captures the evolution of the channel with time.
% In essence, the complex normal distribution of the underwater channel is time-varying, leading to temporal variations in the mean, standard deviation and Rician k-factor for the channel. 
% The probability distribution function of the Rician channel is given by
% \begin{equation}
% f_{X_{\text {rice }}}\left(x_r\right) \sim 2 \frac{(1+k)}{\Omega} x_r e^{-\frac{(1+k)}{\Omega} x_r^2-k} I_0\left(2 \sqrt{\frac{k(1+k)}{\Omega} x_r}\right),
% \end{equation}
% where $k$ is the Rician factor, $I_0I_0$ is the zeroth-order modified Bessel function of the first kind and
% $\mathbb{E}\{X^2_{rice} \} = \Omega$.

\subsection{Simultaneous Wireless Information and Power Transfer for Acoustic-based Underwater Networks}\label{sec:swipt-physics}
Acoustic energy propagates in water as fluctuating pressure waves with a given amplitude, frequency, and phase.
In water, acoustic waves are primarily generated through the piezoelectric effect, which is the deformation of certain materials (such as piezoelectric ceramics) due to the application of an electric field.
By carefully altering the frequency and intensity of the applied electric field, a piezoelectric ceramic material (called a transducer) can be expanded and contracted at a desired frequency, causing information-bearing waves to be generated around the material when immersed in water.
The strength of the acoustic waves generated depends on the magnitude of the electrical input power and the characteristics of the transducer, such as its electrical--acoustic conversion efficiency and the impedance matching between the transducer and seawater~\cite{AppliedUnderwaterAcousticsBjorno2017}.
% The acoustic source level depends on the electrical-acoustic transfer function of the source transducer, which is given by~\cite{SoundPrinciplesUrick1975} $\varepsilon = \frac{P_{rad}}{P_{e}}$,
% % \begin{equation}
% % \varepsilon = \frac{P_{rad}}{P_{e}},
% % \label{eq:conv_efficiency}
% % \end{equation}
% where $P_{e}$ is the electrical input power and $P_{rad}$ is the radiated acoustic power.
The shape of the waveform generated depends on the dimensions of the source transducer (point sources generate cylindrical waves in the far field; compact sources 
% (ones whose dimensions are far less than the wavelength of the generated field) 
generate spherical waves; planar sources generate plane waves, etc.).
Similarly, at the remote receiving end, pulsation of the piezoelectric ceramic material generates an electrical signal that can be used to decoded information encoded in the acoustic waves at the source.
% In theory, all the energy put into the water at the source can be recovered at the receiver after accounting for the channel and acoustic--electrical conversion losses, given by equations (\ref{eq:tl}) and (\ref{eq:conv_efficiency}), respectively.
% encountered between the source and receiver is calculated

% A transducer converts one form of energy to another; they serve as acoustic sources and hydrophones (acoustic receivers) underwater.
% Similar to terrestrial-based antennas, the frequency of the acoustic signal dictates the size of acoustic transducers required; high frequency sounds are generated by compact sources whereas low frequency sounds are generated by larger elements or transducer arrays. 
As highlighted in Section \ref{Intro}, 
% in addition to extracting information encoded in acoustic waves,
electrical energy can also be harvested from the received signal.
A propagating acoustic wave in the $x$-axis generates pressure fluctuations that can be detected by an acoustic hydrophone. 
The fluctuations can be expressed as~\cite{BatterylessIoUTGuida2022}
\begin{equation} p(x) = p_0e^{-\alpha x}e^{j(\omega t-kx)}, 
\end{equation}
where $k=\omega/c$ is the wave number and $\omega$ is the angular frequency.
At the target, a receiver hydrophone containing a piezoelectric transformer converts the mechanical vibrations from the acoustic waves across its terminals into electrical energy.
The receiver sensitivity ($\rho$) expresses the minimum acoustic energy per unit pressure that can be detected by the acoustic hydrophone.
It is given by~\cite{SWIPTAnalysisBereketli2012} $ \rho=20\log_{10}M,$
% \begin{equation}
%     \rho=20\log_{10}M,
% \end{equation}
where $M$ is the sensitivity in V/$\mu$Pa.
The voltage induced across the transducer terminals, $pM$ depends on the acoustic pressure $p$ at that point, and can be expressed as $p=10^{({\gamma}/{20})}, $
% \begin{equation}
%     p=10^{({\gamma}/{20})},
% \end{equation}
where $\gamma$ is the received SNR (Eq. \ref{eq:sonar}).
The induced voltage, $V_{ind}$ at the receiver hydrophone terminals is given by~\cite{SWIPTAnalysisBereketli2012}
\begin{equation}
V_{ind}=\left({10^{{\gamma}/{20}}}\right)\left({10^{{\rho}/{20}}}\right).
\end{equation}
The electrical power available for harvesting, $P_{\nu}$ depends on the impedance matching between the receiver hydrophone and the surrounding seawater.
For a single hydrophone, this can be expressed as 
\begin{equation}
    P_{\nu}=\frac{{V_{ind}^{2} }}{{4R_{p}}},
    \label{eq:harvestable_power}
\end{equation}
where $R_{p}$ is the load resistance required to ensure impedance matching.
If a transducer array is used (such as in~\cite{SWIPTAnalysisBereketli2012}), Eq. (\ref{eq:harvestable_power}) is scaled by the number of array elements as $n{ \frac{{\left({V_{ind}}\right)^{2}}}{ 4R_{p}}}$. 
The total harvestable power, $P$ is given by
\begin{equation}
    P= \eta {\frac{{10^{{\left({\gamma + \rho}\right)}/{10}}}}{4R_{p}}}, 
    \label{eq:harvested_power}
\end{equation}
where $\eta$ is acousto-electric power conversion efficiency.

\vspace{-0.95em}

The SNR, $\gamma_{i}$ between the $i$th node and the AUV can be expressed as
\begin{equation}\label{eq:snrCH}
\gamma_{i}=\frac{P_i / TL_{i, auv}} {N_{0}(f) B},
\end{equation}
where $P_i$ is the transmitting power of the $i$th node, $TL$ is the transmission loss between the node and the AUV, $N_{0}$ is the noise power spectral density, and $B$ is the operating bandwidth.
Similarly, the SNR $\gamma_{auv}$ between the AUV and the surface station (SS) can be expressed as
\begin{equation}\label{eq:snrsink}
\gamma_{auv}=\frac{P_{auv} / TL_{auv, ss}}{N_{0}(f) B},
\end{equation}
where $\gamma_{auv}$ is the transmitting power of the AUV, $TL_{auv, ss}$ is the transmission loss between the AUV and SS, and $N_{0}$ is the noise power spectral density.
% Outage occurs when the received SNR falls below a predefined threshold, $\gamma_{th}$.
% For ease of analysis, we assume the same $\gamma_{th}$ for dcoding data at the AUV and at SS.
% Outage occurs when $\gamma_{i} <\gamma_{th}$, or $\gamma_{i} > \gamma_{th}$ but $\gamma_{auv} < \gamma_{th}$, as defined by the following condition
% \begin{equation}
%     P_{\text{out}} = \Pr(\gamma_{i} < \gamma_{th}) + \Pr(\gamma_{i} > \gamma_{th},  \gamma_{auv} < \gamma_{th}).
% \end{equation}
% When the network is not in outage, the throughput per transmission block can be expressed as
% \begin{align}
%    \mathcal{T}= &\frac{(1 - P_{\text{out}}) R (T/2)}{T},
%    \\= &\frac{(1 - P_{\text{out}}) R}{2}
%    \label{eq:throughput}
% \end{align}
% where $R$ is the data transmission rate and $T$ is the total block transmission time.
The system throughput is given by
\begin{equation}
    \mathcal{T} = B \log_{2} (1 + \gamma_{auv}),
\end{equation}
where $\gamma_{auv} \geq \gamma_{th}$, and $\gamma_{th}$ is the minimum receiver sensitivity.

\subsection{Autonomous Underwater Vehicle Dynamics}
% An AUV submerged in water experiences a force which can be resolved into  a horizontal drag force and a perpendicular lift force~\cite{Seaglider}.
The power consumed by the AUV is a function of the forces acting on it.
% AUV designers must strive to achieve an effective drag to lift ratio (called the load factor) to minimise energy losses.
% Its trajectory is governed by three motion primitives: straight movement, right turn at an angle and right turn at an angle, and its path (known as Dubins path) consists only of arcs and line segments.
% The straight ahead movement consumes the least energy to execute. 
% The power consumed in executing turn motions depend on the turning radius, which is in turn affected by the load factor of the AUV. 
The AUV energy usage can be analysed as follows~\cite{AUVEnergy2020}.
The electrical power $P$ used for motion is given by
\begin{equation}
    P=P_{\mathrm{prop}}+H ={D \times V} + H,
\label{eq:auv-power}
\end{equation}
where $H$ is the hotel load (power consumed by the AUV subsystems less the propulsion system), $P_{\text {prop }}$ is the propulsion power, $D$ is the drag force and $V$ is the AUV velocity.
The drag force is a function of the AUV design (mostly shape), speed and hydrodynamic properties of seawater. It is given by
\begin{equation} \label{eq:2}
D=\frac{1}{2} \frac{C_{\mathrm{D}} A \rho v^{2}}{\beta},
\end{equation}
so that Eq. (\ref{eq:auv-power}) can be expressed as
\begin{equation} 
P_{\mathrm{prop}}=\frac{1}{2} \frac{C_{\mathrm{D}} A \rho v^{3}}{\beta},
\label{eq:auv-prop-power}
\end{equation}
which depends on the drag coefficient, ${C_{\mathrm{D}}}$, area of the AUV, $A$, the density of water, $\rho$ and $\beta$, which is a conversion factor between the mechanical power used for diving and the input electrical power into the motor.
% \begin{equation} \label{eq:3}
% \eta=\frac{P_{m}}{P_{e}}
% \end{equation}
The power used to travel between two points $p,q$ is given by $E=P_{\mathrm{prop}} t_{pq} =P_{\mathrm{prop}} \frac{d_{pq}} {v},$
where $t_{pq}$ is derived from the velocity of the AUV and the distance between the points, $d_{pq}=\sqrt{\left(x_{p, q}-\hat{x}_{p, q}\right)^{2}+\left(y_{p, q}-\hat{y}_{p, q}\right)^{2}}$,
where $x_{p, q}, y_{p, q}$ represent the $x$ and $y$ coordinates of the first point, and $\hat{x}_{p, q}, \hat{y}_{p, q}$ represent the $x$ and $y$ coordinates of the second point.
More in-depth coverage of AUV dynamics and power requirements can be found in~\cite{AUVGuidance, AUVPlatforms}.

\begin{table}[]
\label{tab:para}
\caption{Table of Parameters I}
\begin{tabular}{|l|l|}
\hline
\textbf{Parameter} & \textbf{Value}                    \\ \hline
Network size & $100 \mathrm{~m} \times 100 \mathrm{~m} \times 50 \mathrm{~m}$ \\ \hline
Transmit power ($P_{t}$)      & 170 dB re 1 $\mu$Pa @ 1m \\ \hline
Frequency, $f$                                      & 24 kHz                                 \\ \hline
Nominal sound speed, $c$                          & 1500 m/s                                \\ \hline
$f_{min}$, $f_{max}$                         & 22, 26 kHz                             \\ \hline
End-to-end reliability ($\alpha$) & 0.95                                   \\ \hline
Link failure rate ($p_{i}$) (random)          & 0.05 -- 0.25                           \\ \hline
Wind speed, (w)                                & 10 m/s                                 \\ \hline
Shipping activity factor, (s)                  & 0                                      \\ \hline
Spreading factor, ($k$)      & 1.5                \\ \hline
Relay set, ($K$)      & 6                \\ \hline
Node circuit power, ($P_C$)      & $ \approx 100 \mathrm{~mW}$               \\ \hline
Packet size, ($L_t$)      & 100 bytes              \\ \hline
Noise power, ($NL$)      & $-50 \mathrm{~dB}$                \\ \hline
Target transmission rate, ($R$)      & 1               \\ \hline
Learning rate, $\alpha$     & 0.75             \\ \hline
Discount factor, $\kappa$     & 0.99            \\ \hline
Replay memory size $N$   & 1000            \\ \hline
Exploration factor, $\epsilon$ & 1.0            \\ \hline
$\epsilon_{decay}$    & 0.999            \\ \hline
$\epsilon_{min}$  & 0.001           \\ \hline
Batch size   & 4            \\ \hline
\end{tabular}
\end{table}

% \hl{update table of parameters to include all parameters}

\section{Reinforcement Learning Model}\label{sec:rl-model}
% \hl{Consider parts to move to the intro section  and which (higher level/algorithmic parts) to keep under system model}
RL is a branch of machine learning that involves learning through interaction.
A piece of software called an \emph{agent} takes trial-and-error actions in a given environment in order to learn some desired behaviour.
To quantify how well the agent has learned the desired behaviour, a numerical reward is assigned according to the desirability of each action taken towards achieving the end objective.
If the action taken is the desired one, a high reward is obtained; else, a low or negative reward is obtained.
% % Over time, a piece of software (called an agent) learns which actions lead to high rewards and which ones lead to low rewards.
As a result, a reward signal can be used to completely guide the behaviour of an RL agent in a given environment.
% using only a reward function, since such a function, if appropriately designed, will guide the agent to take optimal actions (actions that return the highest rewards) at all times once it has learnt the environment.
The cumulative or total reward that an agent can receive from an environment is called the \textit{return}.
In addition to learning through interaction (online learning), the agent can also be trained offline to identify patterns in large datasets by leveraging the power of neural networks~\cite{OfflineRLSurveyPrudencio2022}, after which it utilises this prior experience to speed up learning when deployed in real environments.
 % reward function is used to guide an agent to learn intelligent behaviour in complex dynamic environments so that it takes only actions that lead to near-optimal performance of the system.
 In RL, feedback is sometimes delayed, 
 % so it might sometimes be preferable to take actions that yield a low reward at the current instance, if such actions will lead to higher overall returns in the future.
and current actions can have an impact on future states of the system and the rewards due to those states.
 Taking actions that have immediate high rewards is referred to as being \textit{greedy}, in which case the agent \textit{exploits} its current knowledge to maximise the reward it can obtain.
The agent can also forego immediate high rewards and test other actions to evaluate if they could lead to higher future rewards. This is known as \textit{exploration}.
 A trade-off is often made between exploration and exploitation to maximise the expected return in an environment.

An RL problem is often modelled as an MDP, which provides a mathematical framework for modelling decision-making.
MDPs exhibit the Markov property (the future state of a system depends solely on its current state and is conditionally independent of the past), and consists of a tuple $M = (S, A, P, R_{t}, \kappa)$, where $S, A, P, R_{t}$ respectively represent the state space, action space, environment dynamics and transition probabilities and reward function at a time, $t$; $\kappa$ represents a discount factor that indicates how much priority is given to immediate rewards compared to future rewards.
More in-depth coverage of the foundations of RL can be found in~\cite{RLBookSutton2018} while a concise overview of its applications in the IoUTs can be found in~\cite{RLIoUTs}.
In RL terminology, an \textit{agent} is some software trained to perform a given task, such as embedded software in an AUV used for underwater navigation. The agent is represented by its \textit{state}, which is (internal) information from the environment that it uses to select its next action.
% \hl{Concept of return as a measure of long-term reward}
Everything external to the agent constitutes its \textit{environment}, from where it draws sensory inputs. 
% A \textit{state} represents the information that the agent uses to decide the \textit{action} to take in the environment at each time instant.
A \textit{reward} is a numerical feedback signal used to judge if the agent selected the correct action for the given state and environment.
When the agent takes an action or a set of actions in a given state, it transitions to the next state and receives a reward for the previous action.
% A comprehensive overview of RL and its applications in the IoUT domain can be found in~\cite{RLIoUTs}.

In RL, an agent can learn a \emph{policy function}, a \emph{value function} or a \emph{model} of the environment. A policy, $\pi$ is a mapping from states $S$ to actions $A$.
That is, a policy can be defined as the probability of selecting a particular action $a$, in a given state $s$: ($\pi(s,a)=p(A(t)=a\mid S(t)=s$).
% An optimal policy function is a set of rules for selecting actions that if followed, leads to the maximum cumulative reward that can be obtained from the given state.
In \textit{deterministic policies}, every state has a discrete set of actions associated with it, whereas the set of actions is derived from a probability distribution in \textit{stochastic policies}.
\textit{On-policy} learning occurs when the data being generated is used to learn the best policy, changing the behaviour of the agent in the course of learning (e.g. the SARSA algorithm), whereas \textit{off-policy} learning requires two policies: one for taking actions in the environment and the other for learning from the data generated (e.g. policy gradient (PG) methods). A \textit{value function} measures how good a given state is (it quantifies the cumulative reward that can be derived from a given state over time).
It can be a function of a state (state-value functions, $V^\pi(s)$), or of a state-action pair (action-value functions, $Q^\pi(s,a)$).
A \emph{model} or \emph{transition function} is an ensemble of the environmental knowledge available to the agent. It indicates how the agent represents the environment and enables the agent to predict future states and actions, even without actual interaction with the environment. 
% RL Algorithms
An agent using \emph{model-based} RL methods learns the transition dynamics of the environment to select the optimal action to take in every state, whereas in \emph{model-free} methods, it must interact with the environment to learn an optimal policy or value function that maximises the expected return.
% RL algorithms based on this technique are called \textit{model-based} methods.
% If a model is not available, an agent must interact with the environment to learn an optimal policy or value function that maximises the expected return.
% This is the principle behind \textit{model-free} RL methods.
% Model-based methods combine \emph{planning} and \emph{learning} (ref. ~\cite{SurveyModelBasedRLMoerland2020}); in planning, the agent relies on a known model of the environment to plan actions to take, whereas in learning, it uses the transition dynamics of the environment to compute an optimal policy.
% \hl{cite RL magazine for anyone wishing to learn more}

In model-free methods, the agent can directly learn an optimal policy (PG methods) or indirectly learn a value function (value iteration methods), which is then used to extract the underlying policy. Hence, model-free methods can be \textit{value-based} or \textit{policy-based}.
Policy-based algorithms (more details can be found in~\cite{PolicyBasedRLSewak2019}) iteratively optimise a policy through interaction with the environment until it converges to an optimal policy.
Value-based methods optimise a state-value function $V^\pi(s)$ or a state-action value function $Q^\pi(s,a)$,  also called a $Q$ function (e.g.  $Q$-learning, SARSA, multi-armed bandits, etc). 
The state-value function $V^\pi(s)$ indicates the expected return in a given state (a measure of how good the state is) following a policy $\pi$, and is defined as $V^\pi(s)=\mathbb{E}_\pi\left[R_t \mid s_t=s\right]$.
The $Q$ function denotes how good it is to take an action in a given state following a policy $\pi$, and is defined as $Q^\pi(s, a)=\mathbb{E}_\pi\left[R_t \mid s_t=s, a_t=a\right]$.
Once the $V^\pi(s)$ or $Q^\pi(s,a)$ is found, it can be used directly to extract the optimal policy~(argmax of the optimal value function).
The $Q$ function uses a $Q$-table to show all the actions that the agent can take in a given state and their corresponding values, hence the action with the maximum value is selected as the optimal action.
% The drawback is that it requires a lot more data to learn compared to $V^\pi(s)$, and the memory and computational requirements increases rapidly as the number of states and actions grow (temporal difference learning methods can help to address this problem, whereby neural networks are used to generate $Q$ estimates for given ($s,a$) pairs of inputs).
% \hl{Need to provide background for why VB methods were used in this paper}
% Examples of value-based algorithms include $Q$-learning, state-action-reward-state-action~(SARSA), multi-armed bandits (MAB), etc.
% Policy-based algorithms iteratively optimise a policy through interaction with the environment until it converges to an optimal policy~\cite{PolicyBasedRLSewak2019}.
% Policy-based RL approaches are covered in~\cite{PolicyBasedRLSewak2019}.
% Since different policies and actions lead to different expected returns, 
The Bellman optimality equations are used to find optimal policies and value functions.
% The optimal value function is the one with the maximum value (maximum of the $Q$ function) among all the possible value functions, while the optimal policy yields the maximum value function if followed.
For the state-value function, this is given by~\cite{RLBookSutton2018} 
% for a given policy $\pi$, it is defined as 
% \begin{multline}
%   v_\pi(s) 
% =\sum_a \pi(a \mid s) \sum_{s^{\prime}, r} p\left(s^{\prime}, r \mid s, a\right)\left[r+\gamma v_\pi\left(s^{\prime}\right)\right], \quad 
% \\
% \text { for all } s \in \mathcal{S},  
% \end{multline}
% where $p$ is the probability of transitioning from a state $s$ to $s'$ after taking an action $a$, $r$ is the reward associated with $a$ and $\gamma$ is the discount factor.
% The Bellman optimality equations provide analytical proofs of the existence of a maximum value function for a given state, and the optimal action that can result in this value.
% \hl{Bellman equations are outlined below - summarise and cite!}
% \begin{equation}
% % \begin{aligned}
% q_*(s, a)
% =\sum_{s^{\prime}, r} p\left(s^{\prime}, r \mid s, a\right)\left[r+\gamma \max _{a^{\prime}} q_*\left(s^{\prime}, a^{\prime}\right)\right]
% % \end{aligned}
% \end{equation}
\begin{equation}
% V^*(s)=\max _a \sum_{s^{\prime}} \mathcal{P}_{s s^{\prime}}^a\left[\mathcal{R}_{s s^{\prime}}^a+\gamma \sum_{a^{\prime}} Q^\pi\left(s^{\prime}, a^{\prime}\right)\right]
V^*(s) = \max_{a \in A} \left( R(s,a) + \gamma \sum_{s' \in S} P(s'|s,a) V(s') \right).
\end{equation}
% $V^*(s)$ is the optimal state-value function for state $s$, $A$ is the set of possible actions in state $s$, $R(s,a)$, is the reward for taking action $a$ in state $s$, $\gamma$ is the discount factor, $P(s'|s,a)$ is the probability of transitioning to state $s'$ given that action $a$ is taken in state $s$, and $S$ is the set of possible states.
The Bellman optimality equation for state-action value functions is given by
\begin{equation}
  Q^*(s,a) = R(s,a) + \kappa \sum_{s' \in S} P(s'|s,a) \max_{a' \in A} Q(s',a'),
\end{equation}
where $V^*(s)$ is the optimal state-value function for state $s$, $Q^*(s,a)$ is the optimal state-action value function for taking action $a$ in state $s$, $R(s,a)$ is the reward for taking action $a$ in state $s$, $\kappa$ is the discount factor, $P(s'|s,a)$ is the probability of transitioning to state $s'$ given that action $a$ is taken in state $s$, $S$ is the set of possible states, $A$ is the set of possible actions in state $s$, and $\max_{a' \in A} Q^(s',a')$ is the maximum expected future reward over all possible actions in state $s'$.
The Q-learning algorithm satisfies the Bellman equation and is given by
\begin{multline}
     Q(S_{t}, A_{t}) \leftarrow  Q(S_{t}, A_{t}) +\rho\left[R_{t+1}+\kappa \max _a Q(S_{t+1}, a)   - Q(S_{t}, A_{t})\right],
  \label{eq:Qlearning}
\end{multline}
where $\rho$ is the learning rate.
% The solution to the Bellman optimality equation is found through dynamic programming, which breaks a complex problem into smaller sub-problems that are solved individually, reusing solutions where sub-problems reoccur. 
% % \hl{what are all the solution options}
% This iterative process can be applied on the value functions or on policies.
% Value iteration starts with a random value function and finds the Q function for it.
% The policy for the $max Q$ is then the optimal policy.
% Policy iteration starts with a random policy and runs through all the policies, calculating the value function for each until the optimal policy (one with $max V^{\pi}$) is found.

% Check~\cite{DRLAsynchronousWakeUpWUSNsSu2021} for a description of parameters.

\subsection{Joint maximisation of throughput and WPT as an RL problem}
The adoption of RL in this article to address the joint maximisation of throughput and WPT problem is motivated by its suitability in solving decision-making problems~\cite{RLIoUTs} in dynamic environments. 
The underwater channel is stochastic and varies with time~\cite{StatisticalChannelModellingQarabaqi2013, UWAChannelModelingMorozs2020}, making the instantaneous SNR unpredictable.
The problem of resource allocation in such channels naturally lends itself to RL solutions~\cite{MDPsEHFadingChannelsLi2015}, whereby the underlying channel parameters can be learned in an online manner through interaction with the network environment.
In this work, the RL models were trained offline, relying on Monte Carlo simulations to explore extensive variations in the underwater channel.
After training, they are then deployed to the simulated underwater network environment where they execute their offline learning to dynamically take optimal decisions despite the rapid variations in the underwater channel.
The problem of jointly maximising throughput and WPT was modelled as an MDP, in which the AUV represents the agent; the underwater environment represents the environment of the MDP; moving a unit distance in any either the positive or negative direction of the Cartesian coordinates (\textit{x, y, or z)} represents actions; the 3D position of the AUV and residual energies of the nodes under its coverage represent the AUV's state; while the reward is defined as a function of the system throughput and wireless power transferred, as well as the energy expended by the AUV to achieve these tasks.
Detailed analysis of the RL parameters is covered in Section \ref{sec:rl-parameters} while the $Q-$learning algorithm is presented in Algorithm \ref{alg:Q-learning}.
\begin{algorithm}[H]
\caption{$Q-$learning-based trajectory selection for underwater SWIPT}\label{alg:Q-learning}
\begin{algorithmic}[1]
\Require Learning rate $\alpha \in (0,1]$, splitting factor $\Gamma$,
discount factor $\kappa \in (0,1]$, set of actions $\mathcal{A}$, Q values $\mathrm{Q}=0$, 
% optimal policy $\pi=\pi^*$
\State Initialize the environment and get initial state $s_1$
\For{$k=0,1,2, \ldots, K$}
% \State The AUV receives data and transfers power to nodes within its coverage with $\Gamma$
\State Select action $a_t \in \mathcal{A}$ using the $\epsilon$-greedy policy
\State Apply action $a_t$ to the environment
% \If{$a_t=m$}
% \State Transmit information to the node D via the $m$-th relay node
% \ElsIf{$a_t=0$}
% \State Transmit information directly to the node D
% \EndIf
\State Obtain $(\mathcal{G}_{t-1}, \mathcal{L}_{t}, \Xi_{t})$ from the environment
\State  Receive rewards $R(s_t,a_t)$ using Eq.~(\ref{eq:reward})
% \State Update $\pi(s_t)$ using Eq.~(\ref{eq:Qlearning})
\State Set $S_{t+1} = (\mathcal{G}_{t}, \mathcal{L}_{t}, \Xi_{t})$
\State Update $Q(s_t,a_t)$ using Eq.~(\ref{eq:Qlearning})
\EndFor
\end{algorithmic}
\end{algorithm}

\subsection{Reinforcement Learning Parameters}\label{sec:rl-parameters}
    % The two design goals of maximising throughput and harvested power considers the channel between the AUV and the network nodes, and the optimal trajectory of the AUV that maximises the design goals.
    % Therefore, the state space $\mathcal{S} = (\mathcal{G}_{t}, \mathcal{L}_{t}, \Xi_{t}) \in \mathcal{S}$ where $\mathcal{G}_{t}$ represents a vector of the channel gain at $t$ obtained via implicit ARQ at periodic intervals, $\mathcal{L}_{t}$ represents a vector of the $(x,y,z)$ coordinates of the AUV at $t$ and $\Xi_{t}$ represents a vector of the residual energies of the nodes under the coverage of the AUV at position $\mathcal{L}_{t}$.
    % Whereas the 3D location of the nodes can be floating point numbers, in our implementation, $(x,y,z) \in \mathcal{L}_{t}$  are integers to reduce the computational demand of representing the system state.
    % % (memory and computing power -- curse of dimensionality).
    % By accounting for the residual energy in the state equation, the AUV is motivated to move to a new state (new location) that enables it to maximise the reward.

\subsubsection{States}
% \begin{itemize}
%     \item \textbf{\textit{States}}:
    The two design goals of maximising throughput and harvested power considers the channel between the AUV and the network nodes, and the optimal trajectory of the AUV that maximises the design goals.
    Therefore, the state space $\mathcal{S} = (\mathcal{G}_{t}, \mathcal{L}_{t}, \Xi_{t}) \in \mathcal{S}$ where $\mathcal{G}_{t}$ represents a vector of the channel gain at $t$ obtained via implicit ARQ at periodic intervals, $\mathcal{L}_{t}$ represents a vector of the $(x,y,z)$ coordinates of the AUV at $t$ and $\Xi_{t}$ represents a vector of the residual energies of the nodes under the coverage of the AUV at position $\mathcal{L}_{t}$.
    Whereas the 3D location of the nodes can be floating point numbers, in our implementation, $(x,y,z) \in \mathcal{L}_{t}$  are integers to reduce the computational demand of representing the system state.
    % (memory and computing power -- curse of dimensionality).
    By accounting for the residual energy in the state equation, the AUV is motivated to move to a new state (new location) that enables it to maximise the reward.
    
   %  \hl{Equation representation of states -- or state transitions from i--j at least
   % added}
   % In the given environment, if the agent begins in "state start," which is represented by (0,0,0, [0,0,0]), it may either reach "next state +R" of (0,0,1, [0,0, 0]) in which case the reward is R = 0 if the action selected led to "Z" direction movement and there is no node that harvests or transmits data, resulting in an energy level of [0,0,0].

   %  For some reason, if any node is present during the agent's action, the reward will be based on the equation XX, and the next state will be (0,0,1,[0,0,+R]) if only nodes three harvest energy or transmit data based on the split factor $eta$.
    
    % In this setup, there is no terminal state, although the agent stops exploring when the number of actions reaches 50.
    % Nevertheless, during agent movement in the environment, the agent is not permitted to move to state {(5,1,5), [energy level]}from state {(0,0,0), [energy level]} because it is an invalid state and the agent can only move along a single axis.
    % Also, if the agent reaches any of the terminal states or the so-called stop exploring state "T" which is a function of the number of actions, the episode ends with no rewards, unless there are nodes that harvest or transmit data.

\subsubsection{Actions}
    
\begin{figure}[t]
    \centering
    \includegraphics[width=8cm]{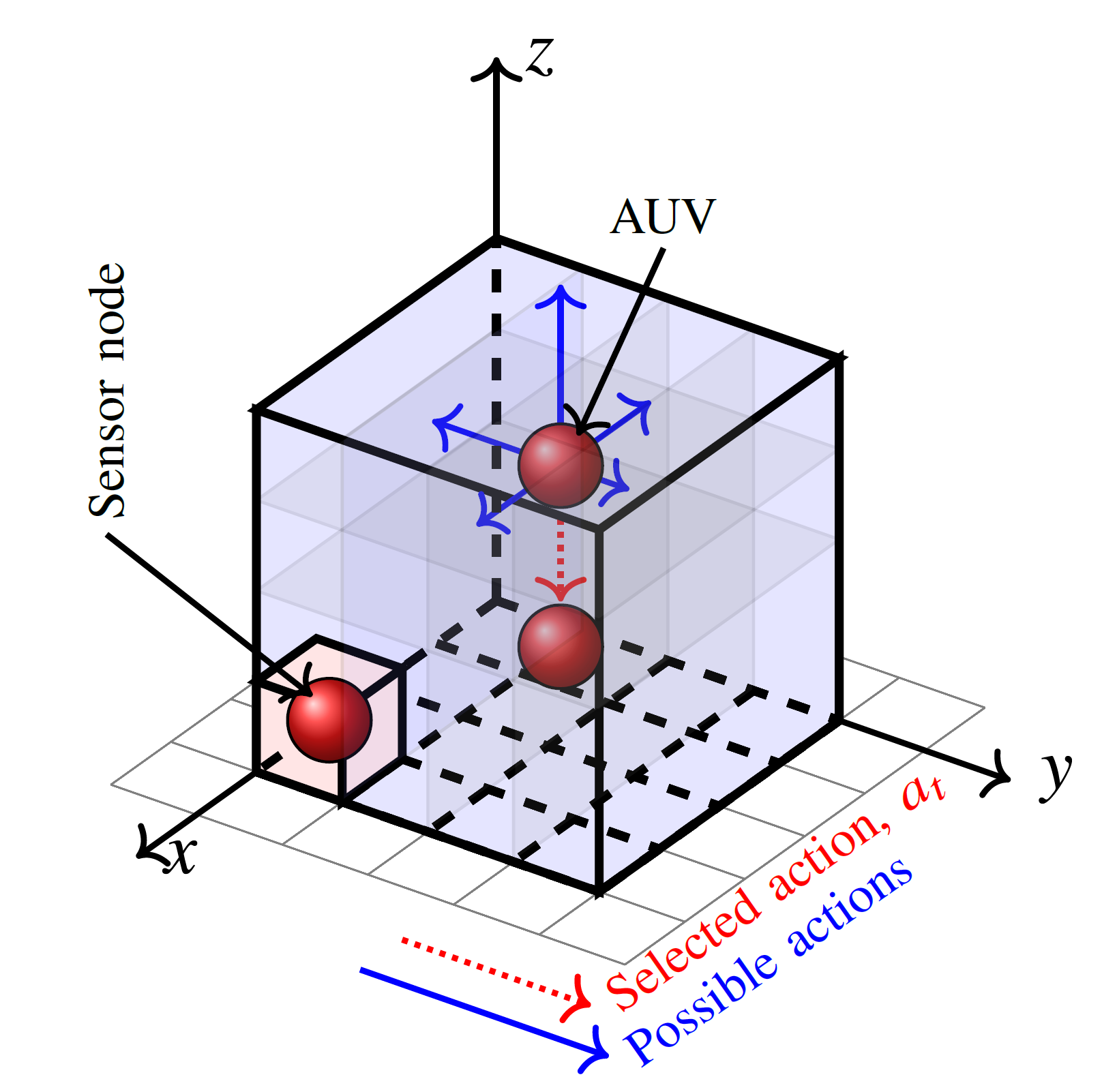}
    \caption{Action space of the AUV agent. Each action involves taking a unit step in any of the $x, y, z$ axes towards a location or set of locations that allows the AUV to maximise the data collection throughput and amount of power harvested, thereby forming a trajectory of points during each episode.}
    \label{fig:action_space}
\end{figure}

    % \item \textbf{\textit{Actions}}: 
    % \hl{Location of the AUV -- e.g. move to next location or stay?}
    The actions consist of taking unit steps towards the location that jointly maximises throughput and harvested power.
    The AUV is considered a rigid body in the 3D underwater space.
    % Each action involves taking a discrete step towards this optimal location.
    The actions are discretised unit steps in the positive or negative $x$ or $y$ or $z$ axes, as shown in Fig.~\ref{fig:action_space}.
    % That is, the agent and environment interaction take place through actions consisting a set of discrete steps given by 
    % \hl{equation} at time steps $t= 1, 2,3, ... n$, where $n$ is the \hl{number of episodes}.
     % \hl{Show an example of state transitions, based on the action selected.}
    % e.g. if in $state_{1} (1 1 1)$ and action $+x$ selected, the AUV (agent) transitions to $state_{1} (2 1 1)$.
    % For state2 (including residual energy), this would be something like: if $a_{s_t,t} = a_{+x}$, $state_{2} (1 1 1 [0 0])$, new state is $state_{2} (2 1 1 [5 0])$, indicating that an energy of 5J was harvested for the action taken, and the appropriate reward accrued.
    Each action is selected following the $\epsilon$-greedy algorithm for both $Q$-learning and SARSA $Q$-learning.
    That is, at each episode, the action is selected based on the following condition   
    \begin{equation}
    a_t = \begin{cases}\underset{a \in \mathcal{A}}{\operatorname{argmax}} Q(s, a), & \text { with probability } 1-\epsilon \\ \sim \mathcal{N}(\mathcal{A}), & \text { with probability } \epsilon,\end{cases}
    \end{equation}
    where $\mathcal{N}$ is the normal distribution that denotes selecting a random action over the set of actions $\mathcal{A}$ and $\epsilon$ is the exploration factor.

    % \begin{figure}[htbp]
    % \centering
    % \includegraphics[width=0.5\columnwidth]{img/actions.png}
    % \caption{The AUV agent takes actions by moving a unit step in either the positive or negative $x$ or $y$ or $z$ axis.}
    % \label{fig:actions}
    % \end{figure}

    % \begin{figure}[htbp]
    % \centering
    % \includegraphics[width=0.5\columnwidth]{img/actions2.png}
    % \caption{The AUV agent takes actions by moving a unit step in either the positive or negative $x$ or $y$ or $z$ axis.}
    % \label{fig:actions}
    % \end{figure}

%     \begin{figure}[t]
% 	\centering
% 	{\resizebox{0.99\columnwidth}{!}{\input{figs/action_space.tikz}}}
% 	\caption{Action space of the AUV agent. Each action involves taking a unit step in any of the $x, y, z$ axes towards a location or set of locations that allows the AUV to maximise the data collection throughput and amount of power harvested, thereby forming a trajectory of points during each episode.}
% 	\label{fig:action_space}
% \end{figure}

\subsubsection{Rewards}
    
    % \item \textbf{\textit{Rewards}}:
Reward design is of paramount importance in RL because the reward controls the agent's behaviour and guides it to take optimal actions.
    % When properly designed, a reward function alone can be used to guide an RL towards taking only optimal actions.
    % Each action transitions the agent from its current state to the next state, and a reward is obtained for the action in the prior state. 
The reward for the state--action--next state interaction is governed by the function $r: \mathcal{S} \times \mathcal{A} \times \mathcal{S} \rightarrow \mathbb{R}$ which we define 
    % the reward for performing an action $a_t \in \mathcal{A}$ in state $s_t \in \mathcal{S}$ 
    as
    \begin{equation}
    \mathcal{R}= \begin{cases}\underset{a \in \mathcal{A}}{\operatorname{\Gamma\mathcal{T} +(1-\Gamma)\mathcal{E}_{\upsilon} - \mathcal{P}_{m}}}, & \text { if nodes} \\ -\mathcal{P}_{m} & \text {otherwise},\end{cases}
    \label{eq:reward}
    \end{equation}
    where $\mathcal{\tau}, \mathcal{E}_{\upsilon}, \mathcal{P}_{m}$ and $\Gamma$ respectively represent the system throughput, harvested power, power consumed to execute a round, and a switching parameter used by the agent to prioritize either achievable throughput or energy harvesting.
    % \hl{$\eta$ ALREADY IN USE}
    % Where $\eta$ represents the switching function that the agent uses o either prioritize the achievable throughput or energy harvesting,
    The parameter $\Gamma$ is defined as $\{\Gamma \in \mathbb{R}: 0 \leq \Gamma \leq 1\}$.
    % While $\eta$ represents the the switching function that agent uses to either prioritilize the the achievable throughput or energy harvested $\{\eta \in \mathbb{R}: 0 \leq \eta \leq 1\}$.
Equation (\ref{eq:reward}) shows that the agent obtains the highest reward by moving to a position that maximises both throughput and harvested power.
However, the agent must take such actions intelligently to conserve its battery, which is also limited in supply. 
The power consumption parameter $\mathcal{P}_{m}$ is used to discourage unprofitable movements.
If it moves to a position with no nodes, it does not receive a reward but is charged a penalty for the energy expended to execute the motion.
% An analytical model of AUV energy consumption for straight and curved motions can be found in~\cite{AUVEnergy2020}. 
% \begin{figure} [t]
% 	\centering
% 	{\resizebox{0.9\columnwidth}{!}{\input{figs/reward_space1.tikz}}}
% 	\caption{Probability that the agent sees at least one node as a function of starting position in 3D environment --- $z=0$.}
% 	\label{fig:prothroughput}
% \end{figure}

\begin{figure}[t]
    \centering
    \includegraphics[width=8cm]{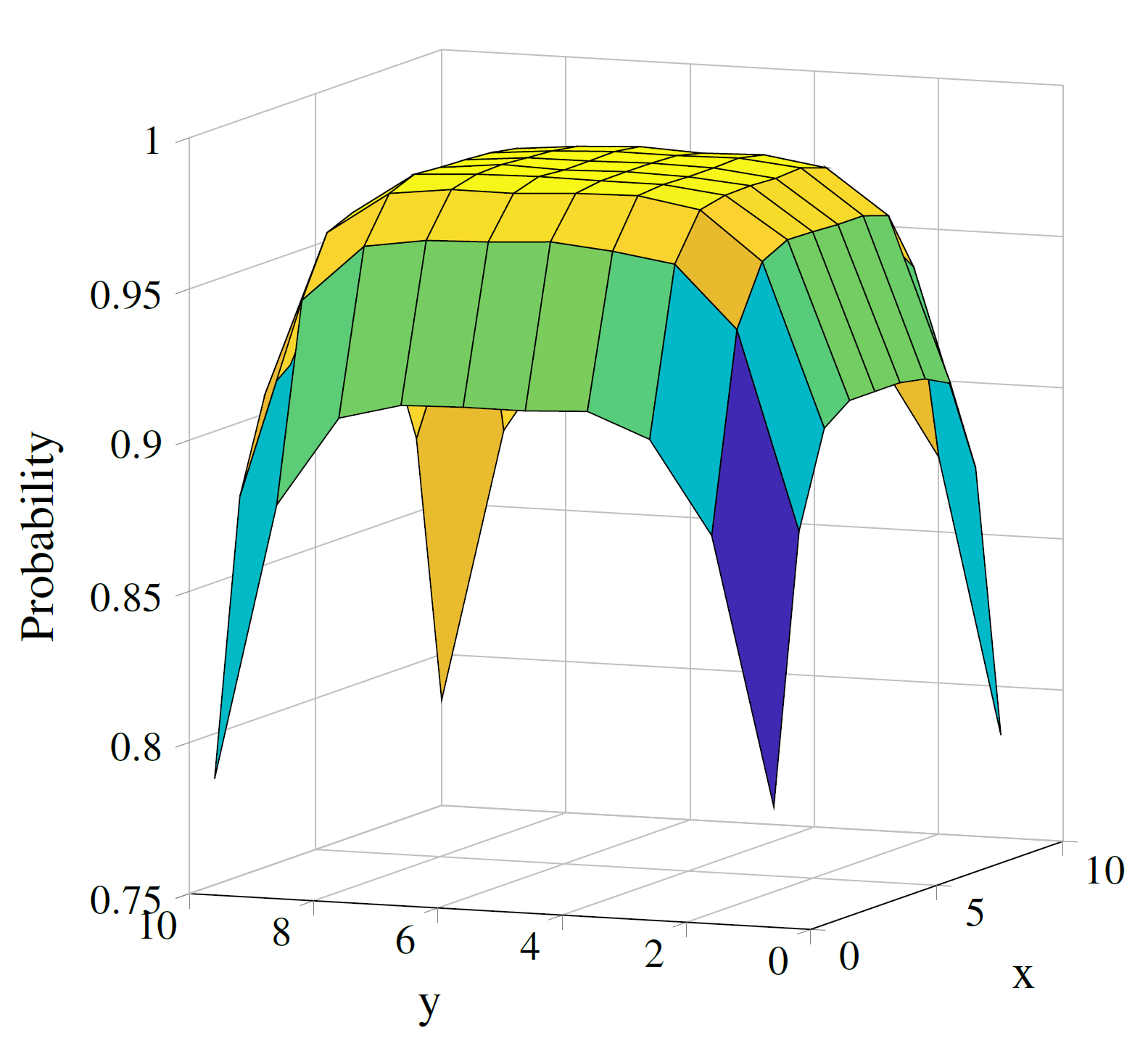}
    \caption{Probability that the agent sees at least one node as a function of starting position in 3D environment --- $z=0$.}
    \label{fig:prothroughput}
\end{figure}

We analysed how the starting position of the AUV impacts the number of underwater nodes it can simultaneously provide coverage to, as shown in Fig.~\ref{fig:model}. 
It should be noted that the starting position is always along the $z$ axis.
Our simulation results show that the probability of covering more than a single node is higher when the AUV is located near the centre of the network area, compared to when it is farther away, as shown in Fig.~\ref{fig:prothroughput}.
To prevent the AUV from becoming stuck in a position with a high probability of covering a large number of nodes (due to the potential to collect more rewards there, at the expense of nodes in sparsely populated parts of the network), we also analysed the probability of covering a specific number of nodes (from 0 to M) as a function of the starting position of the AUV. 
By considering these probabilities, we designed a reward function that encourages the AUV to explore and collect more rewards rather than staying in a position with a high probability of covering many nodes. 
We adopted the binomial distribution as an approximation of the Poisson distribution for our analysis. 
Let $n$ be the total number of nodes in the 3D cube, $x_i, y_i, z_i$ the coordinates of node $i$, $V_{cone}$ denote the volume of the AUV coverage cone with angle $\Theta$ degrees that represent its field of view. 
To calculate $V_c$, we can consider a cross-section of the cone along the plane perpendicular to its axis. 
The cross-section is a circle with radius $r$, where $r$ is the distance between the apex of the cone and the point on the circumference of the circle that is farthest from the apex, and $h$ is the distance from the apex to the base. The volume $V_{cone}$, given by Eq.~\ref{eqn:vcone} always depends on the current position of the AUV, hence, the Monte-Carlo technique can be used to approximate it for the general case as follows
\begin{equation}\label{eqn:vcone}
    V_{cone} = \frac{1}{3} \pi r^2 h.
\end{equation}
Let $p_k$ be the probability of covering $k$ nodes within the cone. 
Since the nodes are distributed randomly in the 3D cube following a Poisson point process with density $\lambda$, the probability of a node being within the cone is given by $p = \frac{V_{cone}}{V_{cuve}}$, where $V_{cube}$ is the volume of the 3D cube.
Considering binomial distribution, the probability of not covering any nodes within the cone is given by $p_0 = (1 - p)^n$.
Similarly, the probability of covering exactly $k$ nodes within the cone can be calculated using the binomial distribution as $p_k = \binom{n}{k} \times p^k \times (1-p)^{n-k}$. 
Therefore, to calculate the probability of covering $k$ or more nodes within the cone, the probabilities of covering $k$, $k+1$, $\dots$, $n$ nodes can be summed up as
\begin{equation}
p_k = \sum_{i=k}^{n} \binom{n}{i} \times p^i \times (1-p)^{n-i}.
\end{equation}

% \begin{figure}\label{fig:expprobnodes}
% 	\centering
% 	{\resizebox{0.9\columnwidth}{!}{\input{figs/reward_space.tikz}}}
% 	\caption{Probability of covering a Specific Number of Nodes Within Field of View as a Function of Initial State Position}
%  \label{fig:proprothroughput}
% \end{figure}

\begin{figure}[t]
    \centering
    \includegraphics[width=8cm]{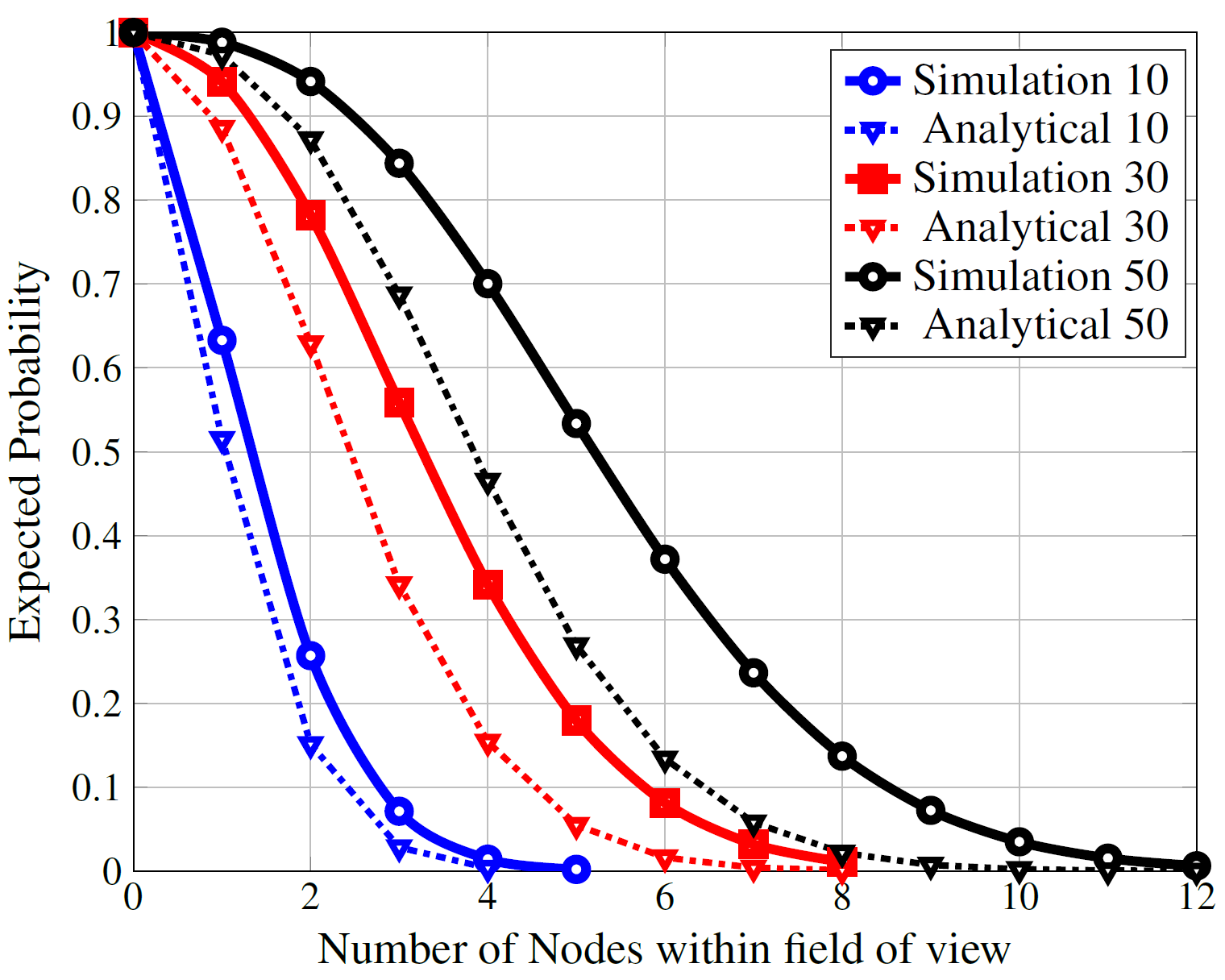}
    \caption{Probability of covering a specific number of nodes within the field of view as a function of initial position of the AUV}
    \label{fig:proprothroughput}
\end{figure}

% \hl{this}
% Figure~\ref{fig:proprothroughput} shows the results from simulation and analytical approach. It can be seen that there is high probability chances greater than 50 $\%$ of covering at least single node when the number of node in the environment is greater then 10, while for number of nodes greater than 30 there are at least high chances of covering more than 3 nodes in staring position and more than 4 nodes when the number of nodes is greater then 50. Hence this distribution might hurt the learning since there will be high chance the learning will stuck in sub-optimal position as number of nodes increase. Hence We add the another penalty of staying in the single position when there is no data to transfer for the case of transmission. So each node will assume to have 100Kb to transmission and as the AUV pass through the node if there is data left the reward agent consider be similar to that in Eqn~\ref{eq:reward} otherwise   the reward is $-\mathcal{P}_{m}$. The simialr approach is considered when energy is harvest to the nodes from the AUV. If the AUV is full charged then the agent get $- \mathcal{P}_{m}$ as reward so to encourage to charge other nodes. 

% \hl{or this}
Figure~\ref{fig:proprothroughput} shows the results from both simulation and the analytical approach. 
The probability of covering at least $M$ nodes is shown to increase as the number of nodes in the environment increases as a function of the initial state (starting position). 
When there are more than 10 nodes in the network, the probability of covering at least one node is greater than 50$\%$. 
For yet higher number of nodes ($>30$), there is a high probability of covering more than 3 nodes in the starting position and an even higher probability ($>50 \%$) of covering more than 4 nodes when the number of nodes exceeds 50. 
This distribution may hinder learning since the agent may become stuck in a sub-optimal position as the number of nodes increases. 
To mitigate this issue, we implemented a penalty for staying in a single position when there is no data to transfer. 
If the AUV passes through a node and there is still data left to transmit, the reward is similar to that given in Eq.~(\ref{eq:reward}). 
Otherwise, the reward is $-\mathcal{P}_m$. A similar approach is taken when the AUV is transferring power to the nodes.
If the node is already fully charged, it receives a reward of $-\mathcal{P}_m$ to encourage it to charge other nodes.

\subsubsection{Environment dynamics}
    % \item \textbf{\textit{Environment dynamics}}:
    Nodes are randomly distributed in a cube of dimensions $L \times W \times H$, where $L$ and $W$ represent the length and width of the network respectively, and $H$ represents the depth below sea level.
    When the agent takes an action $a_{t}$ in state $s_{t}$ given by $(\mathcal{G}_{t}, \mathcal{L}_{t}, \Xi_{t})$, it receives a reward $r_{t}$, and transitions into a new state $(\mathcal{G}_{t+1}, \mathcal{L}_{t+1}, \Xi_{t+1})$.
% \end{itemize}

\section{Performance Evaluation}\label{sec:simulations}
\begin{figure}[t]
    \centering
    \includegraphics[width=8cm]{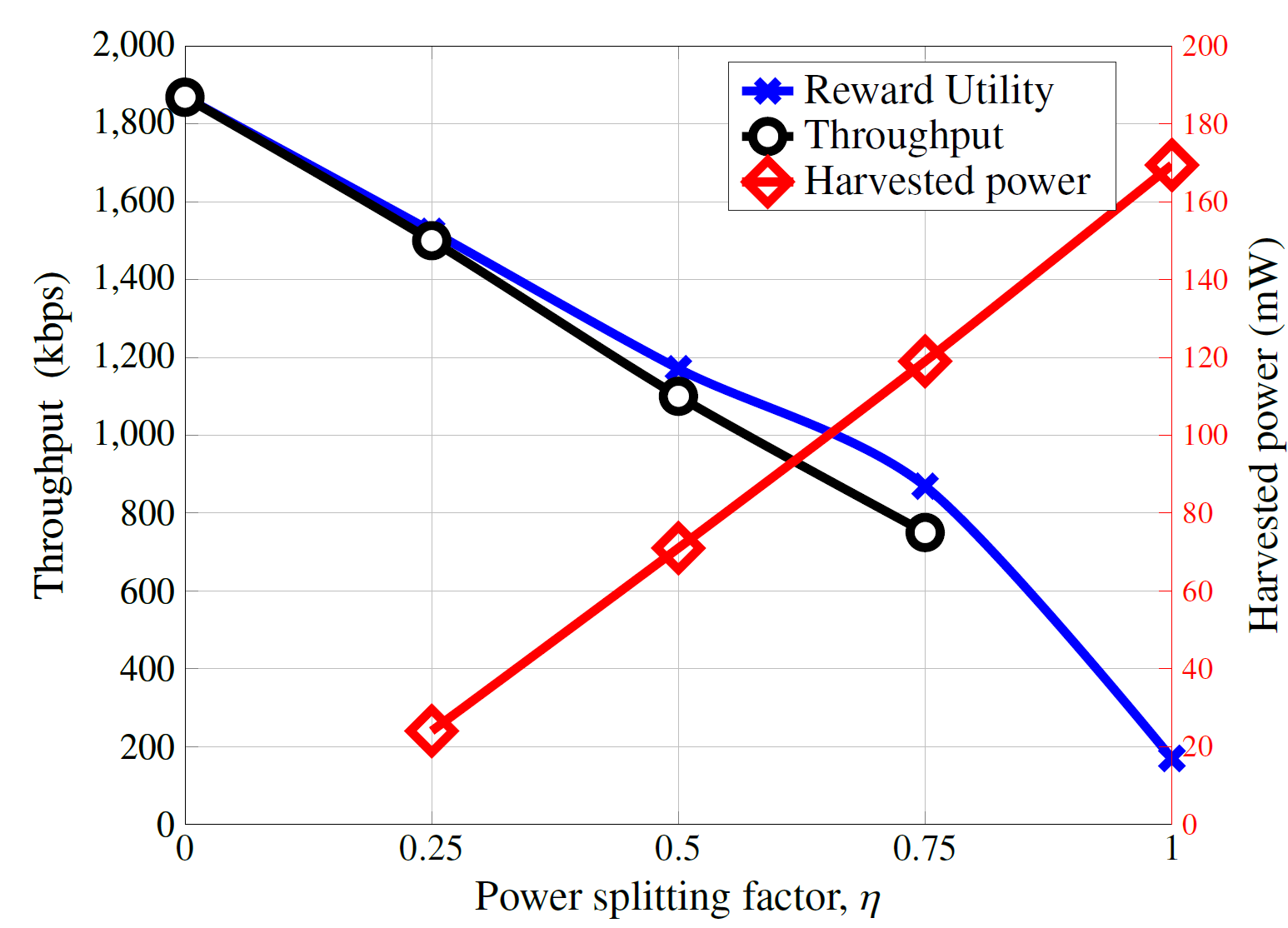}
    \caption{Reward utility for varying values of the PS factor $\eta$ for a network with 25 nodes. It is clear that throughput maximisation dominates irrespective of the splitting factor (as long as $\eta <$ 1) due to the slow charging of the power source compared to the duration required for data transmission.}
    \label{fig:fraction}
\end{figure}

The simulations considered SARSA and Q-learning algorithms based on the $\epsilon-$greedy policy. 
In SARSA, action values are learned by following the current policy, while in Q-Learning, they are learned by following the greedy approach. 
As a result, they converge to the absolute value function under common conditions but at different rates.
Q-Learning tends to converge more slowly compared to SARSA. However, it can continue learning while policies are changed. Additionally, convergence is not guaranteed when Q-learning is combined with linear approximation.
In this study, we examined the performance of the proposed model as the splitting factor, $\eta$ is varied from 0 to 1.
The splitting factor measures how the model allocates resources between two competing objectives, such as WPT and throughput, especially in dense networks.
% We are particularly interested in the scenario where there is a high number of nodes, in the network. 
The performance of the model in this scenario is shown in Fig.~\ref{fig:fraction}, which illustrates the contribution of the utility function to the overall reward.
The model was trained offline to reduce computational demands on the resource-constrained underwater network.
To ensure that the model generalises well to the different network environments, Monte-Carlo simulations were run and the mean results were obtained.

% \begin{figure}[t]
% 	\centering
% 	{\resizebox{0.9\columnwidth}{!}{\input{figs/fraction_harv.tikz}}}
% 	\caption{Reward utility for varying values of the PS factor $\eta$ for a network with 25 nodes. It is clear that throughput maximisation dominates irrespective of the splitting factor (as long as $\eta <$ 1) due to the slow charging of the power source compared to the duration required for data transmission.}
% 	\label{fig:fraction}
% \end{figure}

% In Fig.~\ref{fig:fraction} we investigate the performance of the proposed model in the scenario where we vary the split factor $\eta$ varied from 0 to 1. We only shows only results for scenario where there is high number of nodes. The figure shows the utility function contribution to the overall reward and it can be seen even at value of $\eta=$ 0.5  there is no equal reward contribution even in the convergence since it need the action to stay longer in a given state to maximize the energy harvesting. The rest of the results only show the findings case for the value of  $\eta=$ 1 which maximize the Energy Harvesting and when $\eta=$ 0 which maximize the throughput achieved unless states otherwise.

From the figure, it is clear that even when the splitting factor is set to $\eta=0.5$, the reward contribution is not equal for throughput and harvested power when the learning converges.
This is because the agent must stay longer in a given state to be able to maximise harvested power, which can result in a negative reward for throughput since throughput requires a shorter time to collect the rewards due to a particular state.
This is because it takes a much shorter duration to transfer data from the nodes to the AUV than it takes to recharge the supercapacitor bank of the nodes.
Due to this behaviour, the remainder of the performance evaluations considers only values of $\eta=1$ and $\eta=0$, to respectively maximise harvested power and throughput, unless otherwise stated. These values represent the extreme ends of the reward function and provide valuable insights into the trade-offs between the two design objectives.
% The aim of this simulation was to optimise the trajectory of the AUV. That is, our intent was to find out how the AUV should move from a randomly chosen location at the top of the network in order to maximise the throughput and harvested power.
In terms of the AUV trajectory, we observed that if the agent is constrained to start from a fixed location, a higher number of steps is required before converging to the optimal trajectory. 
This is because the agent will explore more (including taking sub-optimal trajectories). Meanwhile, if the agent starts randomly, it converges faster since fewer steps are required to \emph{orientate} the agent, and it finds the optimal path that it had already learned during training.

% The network was trained offline (to reduce computational demands).
% In order to produce results that generalise to the problem, we considered Monte-Carlo simulation to find the mean solution for different use cases.
% We show the number of steps required to achieve a given throughput.
% % To produce reproducible results, 
% During execution, we evaluate the number of actions required to achieve the maximum reward.
% During training, the agent starts randomly in order to explore more and generalise the environment.
% After training, since the agent has learnt the environment, it starts execution from a fixed location.

% to find the mean solution for different use cases.
% We show the number of steps required to achieve a target throughput.
% To produce reproducible results, 

During execution, we evaluated the number of actions required to achieve the maximum reward.
% The main contribution of this article is the joint optimisation of throughput and harvested power -- by optimising throughput subject to an energy constraint at each sensor node (a minimum harvested power threshold is required to sense and transmit collected data).
% Second, we evaluate the magnitude of harvested power from the AUV.
From a network design perspective, the maximum reward can help to estimate the required power budget for a target throughput.
By setting an energy constraint at the nodes so that they can transmit only if their residual power level is higher than a set threshold, the AUV is forced to maximise its actions to ensure that the target throughput is reached by supplying the power required by the nodes through WPT.
% However, instead of just evaluating Eq. (\ref{eq:harvested_power}) using the proposed model, we present the analysis based on wireless communication metrics.
% In effect, we quantify the power budget required to achieve a target throughput.
This can then be used to quantify the value of electrical power required from the AUV (given the network topology and channel conditions) to supply the target power budget.
This is important for two reasons: a classic challenge for mobile robots used in RL is balancing between continuing to perform assigned tasks (in order to collect higher rewards) versus returning to the charging station to recharge its battery. 
If the robot continues working in the environment, it will collect higher rewards. 
However, if it runs out of power before returning to recharge, it will incur a very high penalty, far higher than any additional rewards it might collect from working longer.
Thus, by quantifying the power budget required to satisfy a target throughput and linking it to the residual energy of the AUV, we successfully address this dilemma for AUVs, ensuring that the AUV knows exactly when to return to the surface station for recharging by monitoring its own battery level.
Secondly, this analysis makes it easy for system designers to prepare a power budget for the network and quantify the expected performance. 
It answers the question: how much power is required to run an underwater sensor network (given the network distribution, reporting rate, density, etc.) and how long will the network last?
The following evaluations show how different models perform under the given network conditions.

\begin{figure}[t]
    \centering
    \includegraphics[width=8cm]{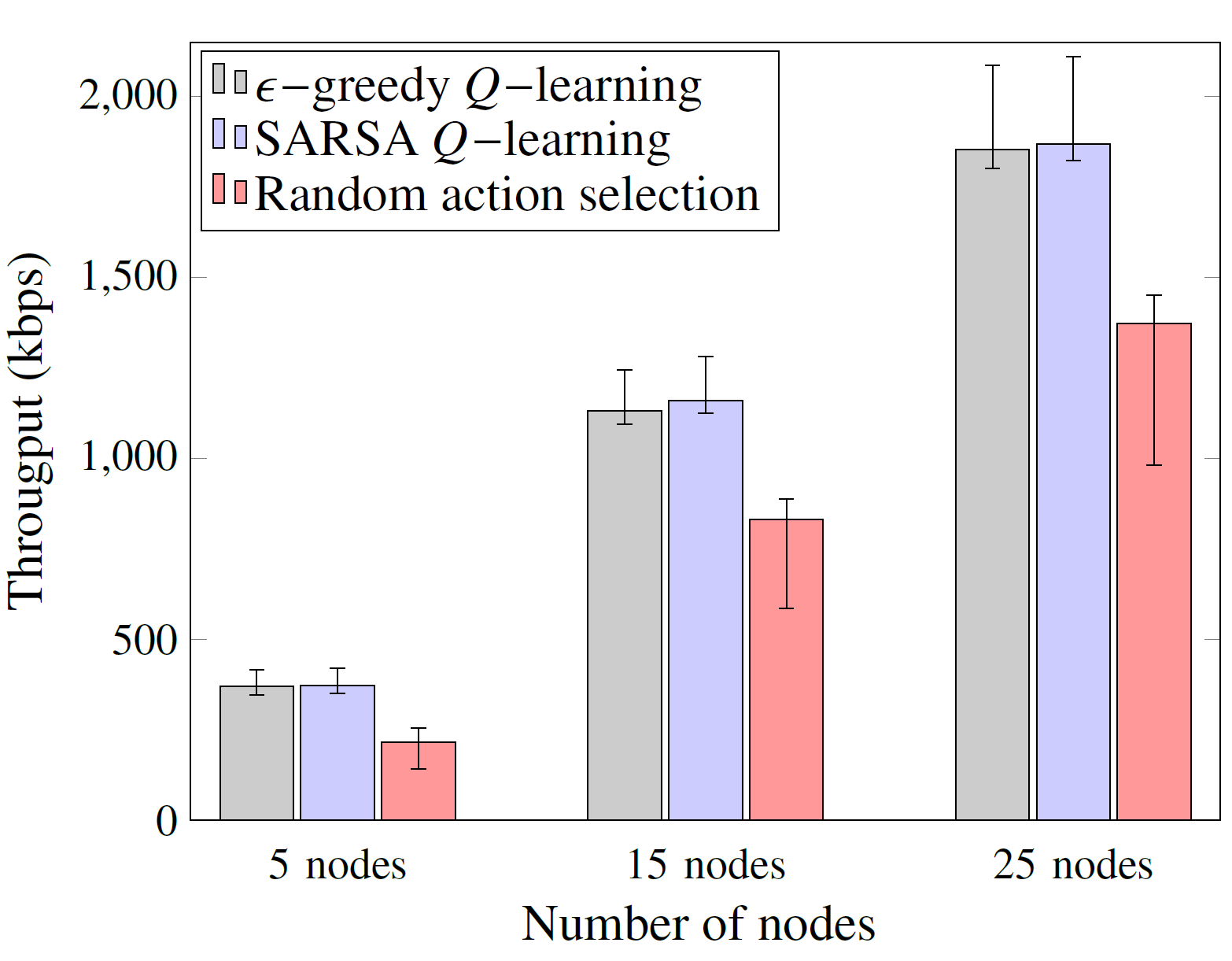}
    \caption{Total network throughput for different algorithms averaged over Monte Carlo runs for different network sizes.}
    \label{fig:throughput}
\end{figure}

% \hl{Repeated discussion of results?}
Most RL models use the average reward per iteration as a metric to evaluate model convergence, with the average reward per episode evaluated and compared with the baseline or other models.
% In Fig.~\ref{fig:throughput}, the average throughput is plotted as a function of the number of iterations.
% Fig.~\ref{fig:throughput} shows the average rewards achieved for throughput for different models.
% Q-learning and SARSA clearly outperform the random scheme. 
% The agent's random actions throughout the trajectory calculate the average throughput in a random scheme. In Fig.~\ref{fig:throughput}, the performance gap between the Q-learning and SARSAQ and the $\epsilon-$greedy action-based scheme is reduced as the number of iterations increases, while random action shows no indication of improvements.
Fig. \ref{fig:throughput} shows the average aggregated network throughput as a function based on the number of actions the agent is required to take over an entire trajectory for different models.
% The throughput is obtained by evaluating the agent's random actions throughout over a trajectory in the random scheme (\hl{clarity required}).
For some applications, $Q-$learning and SARSA yield different results.
% ~\hl{reference for this -- check Sutton}. 
Therefore, we explored how both methods perform under the given optimisation problem.
Fig. \ref{fig:throughput} shows that $Q-$learning and SARSA achieve similar performance but random action selection performs poorly in comparison.
It was observed that the performance gap between the Q-learning and SARSA decreases as the number of nodes increases, but there is no improvement in the random action selection algorithm.
This is because as the number of nodes increases, the spatial variance between the nodes increases, which implies that the agent requires to take more actions (more angles, turns, and steps) to reach all the nodes for data collection, thereby degrading its reward.
If the network size increases further, SARSA begins to outperform $Q-$learning.
Overall, the agent can reach more nodes to collect data as the number of nodes increases, leading to higher total throughput (and hence higher reward).
However, the optimal algorithms (Q-learning and SARSA) return higher rewards than the random action selection.
Each bin in Fig. \ref{fig:throughput} represents the mean aggregate throughput, while the error bars indicate the maximum and minimum throughputs.

% \begin{figure}[t]
% 	\centering
% 	{\resizebox{0.9\columnwidth}{!}{\input{figs/average_action.tikz}}}
% 	\caption{Average number of actions required to achieve a given throughput. The throughput targets are shown in Fig. \ref{fig:throughput}. Clearly, randomly moving the AUV around the network requires far more steps to achieve a target throughput than $Q-$learning and SARSA.}
% 	\label{fig:running_time_thruput}
% \end{figure}

\begin{figure}[t]
    \centering
    \includegraphics[width=8cm]{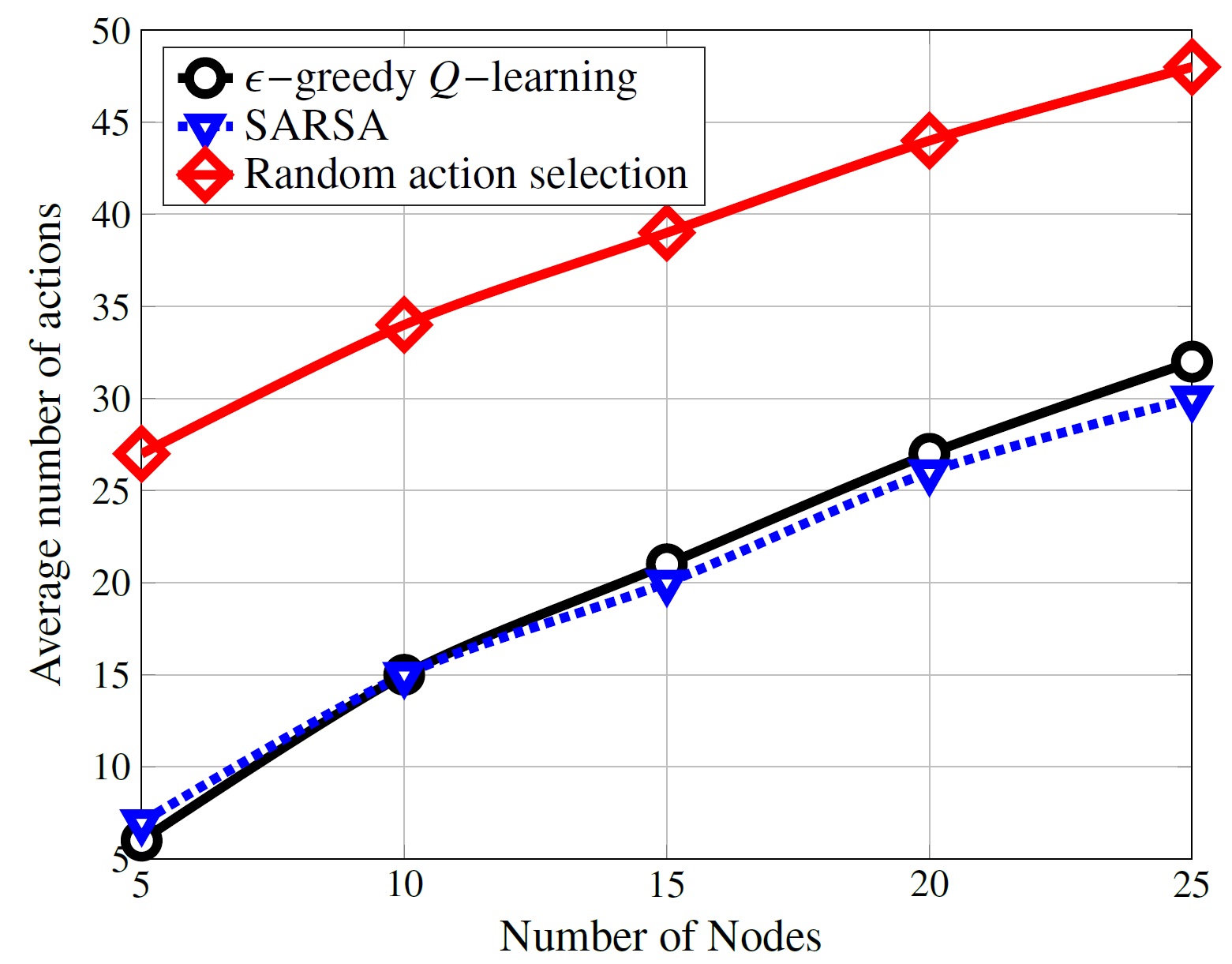}
    \caption{Average number of actions required to achieve a given throughput. The throughput targets are shown in Fig. \ref{fig:throughput}. Clearly, randomly moving the AUV around the network requires far more steps to achieve a target throughput than $Q-$learning and SARSA.}
    \label{fig:running_time_thruput}
\end{figure}

Recall that AUVs are also battery-powered and have a finite supply of energy.
Though they can return to the surface station to recharge their batteries, this affects their operation and might add additional delays to the data collected and limit their usefulness if only one AUV is used.
To improve the endurance of the AUV, it is important that it maximises its energy efficiency by taking optimal actions at all times.
Each action taken by the AUV consumes energy, as shown in Eq. \ref{eq:auv-prop-power}.
Fig.~\ref{fig:running_time_thruput} quantifies the energy efficiency of the AUV in terms of the number of actions it takes to achieve the maximum reward in terms of throughput for different algorithms.
Since the AUV must provide energy to the underwater network nodes via SWIPT and also upload data gathered by the nodes, the energy usage of the AUV takes critical significance in the network deployment.
% To test how the different algorithms perform in terms of preserving the energy efficiency of the AUV, we evaluate the number of actions required to achieve a given throughput.
% Fig. \ref{fig:running_time_thruput} shows the number of actions required to achieve a given average reward (plotted in).
% The result in Fig. \ref{fig:running_time_thruput} compares how the optimal algorithms perform against random actions in terms of the number of actions.
% In addition achieving worse throughput than $Q-$learning and SARSA, it can be seen that random action selection takes far higher number of actions.
Fig.~\ref{fig:running_time_thruput} shows that the optimal algorithms far outperform the random actions by yielding higher throughputs while using much less energy. 
That is, for the same energy usage, $Q-$learning and SARSA yield higher throughput and harvested power.
Stated differently, for the same number of actions, they require less energy to achieve a target throughput, and there is some probability that the random action selection will never achieve the target throughput even if it takes an unlimited number of steps (actions).

% \begin{figure}[th]
% 	\centering
% 	{\resizebox{0.9\columnwidth}{!}{\input{figs/EE.tikz}}}
% 	\caption{Network energy efficiency, which quantifies the power consumed (by the AUV and the sensor nodes) per unit of data collected as a function of the number of nodes for different algorithms.}
% 	\label{fig:EE}
% \end{figure}

\begin{figure}[t]
    \centering
    \includegraphics[width=8cm]{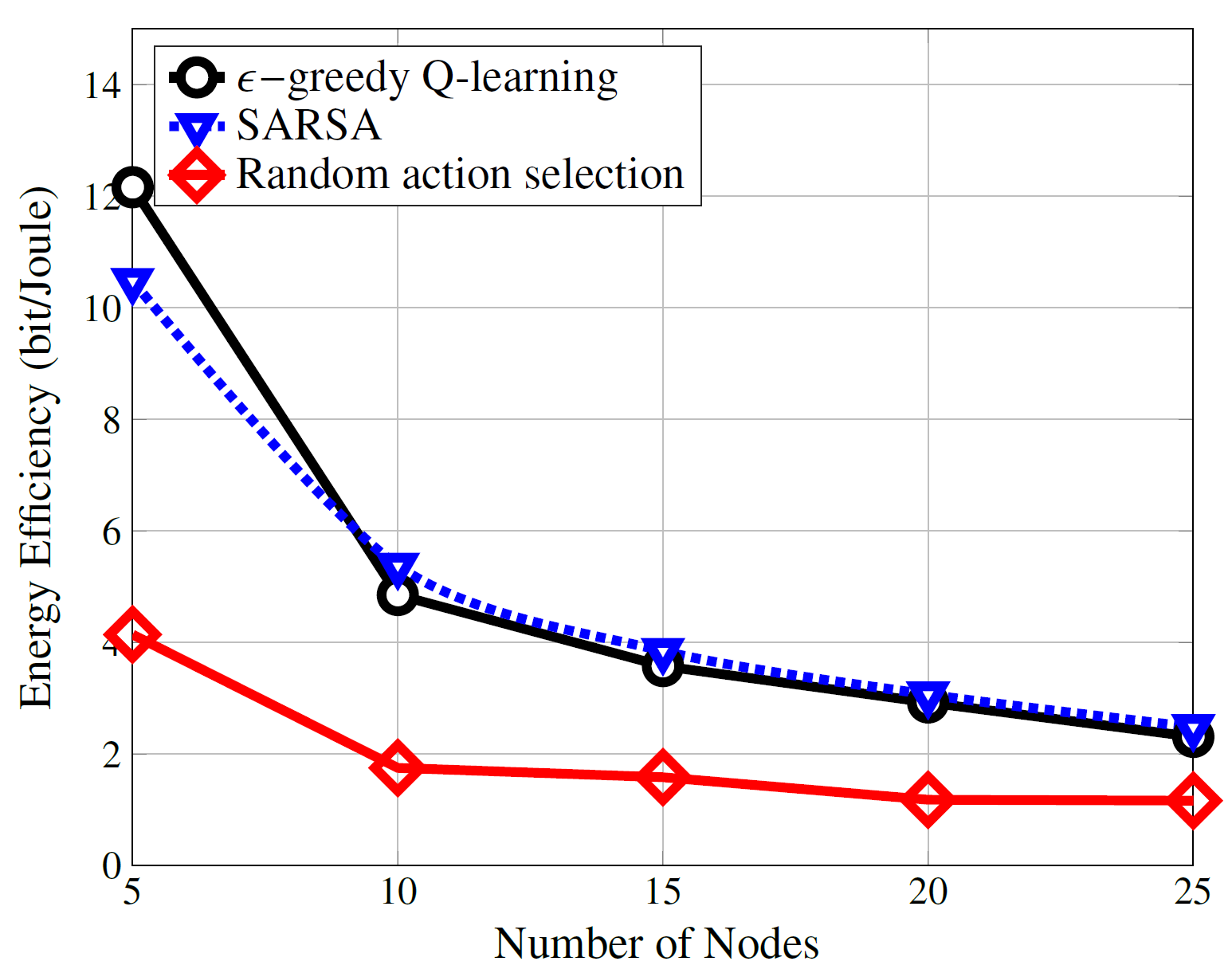}
    \caption{Network energy efficiency, which quantifies the power consumed (by the AUV and the sensor nodes) per unit of data collected as a function of the number of nodes for different algorithms.}
    \label{fig:EE}
\end{figure}

% \begin{figure}[th]
% 	\centering
% 	{\resizebox{0.9\columnwidth}{!}{\input{figs/EH.tikz}}}
% \caption{Total harvested power for different number of nodes per episode (50 steps). This result compares the quantity of harvested power for different algorithms. It is evident that $Q-$learning and SARSA lead to higher harvested power
% than taking random actions. In addition, the optimised algorithms achieve the maximum harvestable power in fewer steps, as shown in Fig. \ref{fig:running_time_power}.
% When the AUV moves (takes actions) at random, not only does it lead to lower harvested power, it might never achieve the maximum power possible in simulation time.}
% 	\label{fig:harvest_power}
% \end{figure}

\begin{figure}[t]
    \centering
    \includegraphics[width=8cm]{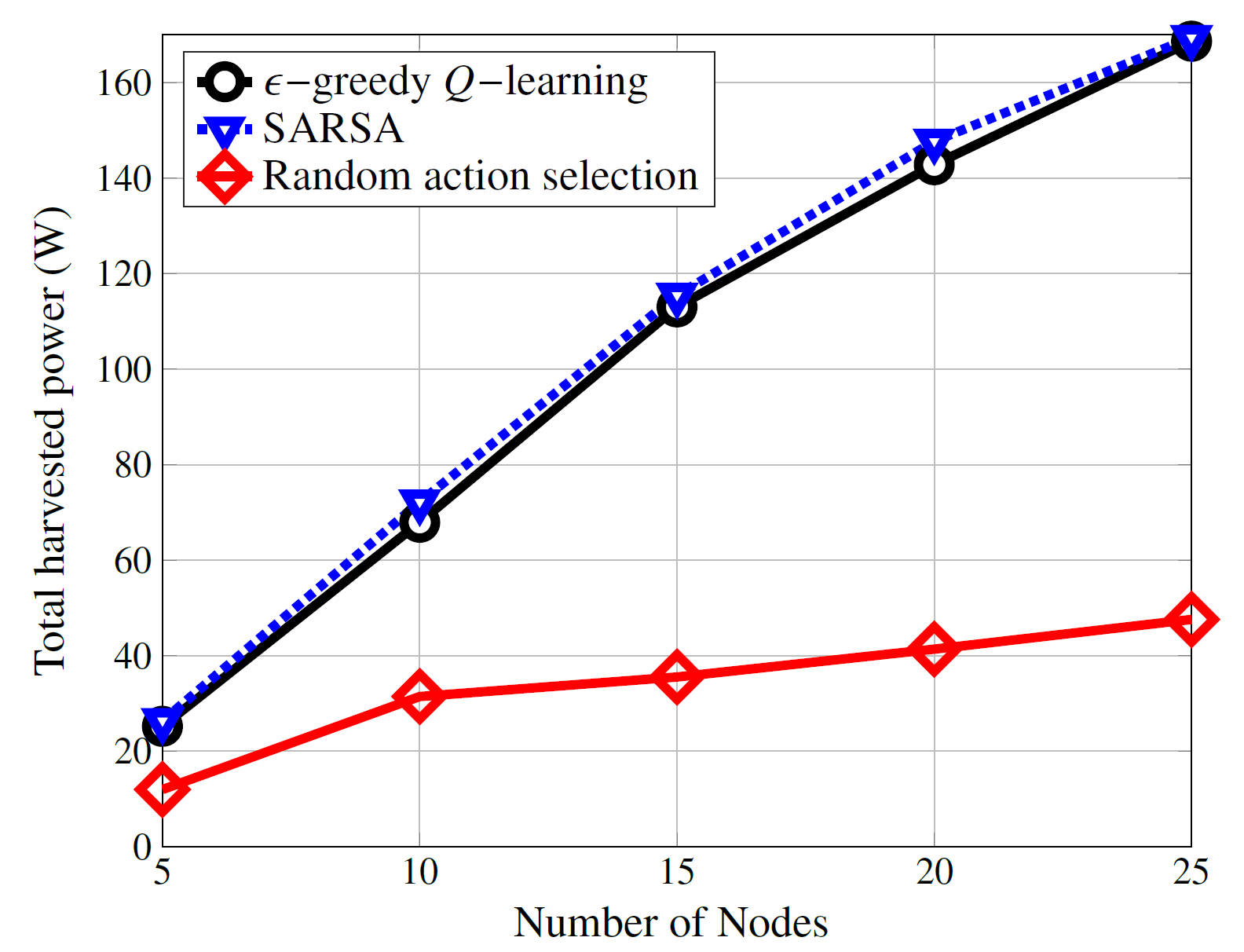}
    \caption{Total harvested power for different number of nodes per episode (50 steps). This result compares the quantity of harvested power for different algorithms. It is evident that $Q-$learning and SARSA lead to higher harvested power
than taking random actions. In addition, the optimised algorithms achieve the maximum harvestable power in fewer steps, as shown in Fig. \ref{fig:running_time_power}.
When the AUV moves (takes actions) at random, not only does it lead to lower harvested power, it might never achieve the maximum power possible in simulation time.}
    \label{fig:harvest_power}
\end{figure}

Energy efficiency can be defined as, $EE = \mathcal{T}/\mathcal{P}_{\tau}$, where $\mathcal{P}_{\tau}$ is the total power.
There are two types of power considered: a) power used by the AUV for navigation, and b) transmission power used for SWIPT.
Each node transmits with 1 kW of power. 
% The total power, $\mathcal{P}_{\tau}$ is the entropy of the system.
$\mathcal{P}_{\tau}$ is the sum of the transmission power from the AUV to node and the navigation power used for propulsion.
The total power for navigation is given by the number of actions per step multiplied by the power consumed by each step.
Fig.~\ref{fig:EE} shows a decreasing trend for energy efficiency because while throughput increases linearly with the number of nodes in the network, power consumption increases geometrically.
However, it is still observed that $Q-$learning and SARSA achieve higher energy efficiency than random movements.

% \hl{Result showing/justifying joint optimisation -- when $\alpha=0.5$. To be extracted from the simulation -- Michael}.

% Figs. \ref{fig:EH} and \ref{fig:running_time_power} indicate that the agent can learn the optimal trajectory that leads to higher harvested power through interaction with the underwater environment, which implies that it can achieve the optimal (target) power harvesting in fewer simulation rounds.
% \begin{figure}[h]
% 	\centering
% 	{\resizebox{0.9\columnwidth}{!}{\input{figs/average_action_PH.tikz}}}
% 	\caption{Average number of actions required to achieve a given magnitude of power harvesting. Clearly, randomly moving the AUV around the network requires far more steps to achieve a target throughput than $Q-$learning and SARSA. Moreover, the agent might never attain the maximum harvestable power when it takes random actions. }
% 	\label{fig:running_time_power}
% \end{figure}

\begin{figure}[t]
    \centering
    \includegraphics[width=8cm]{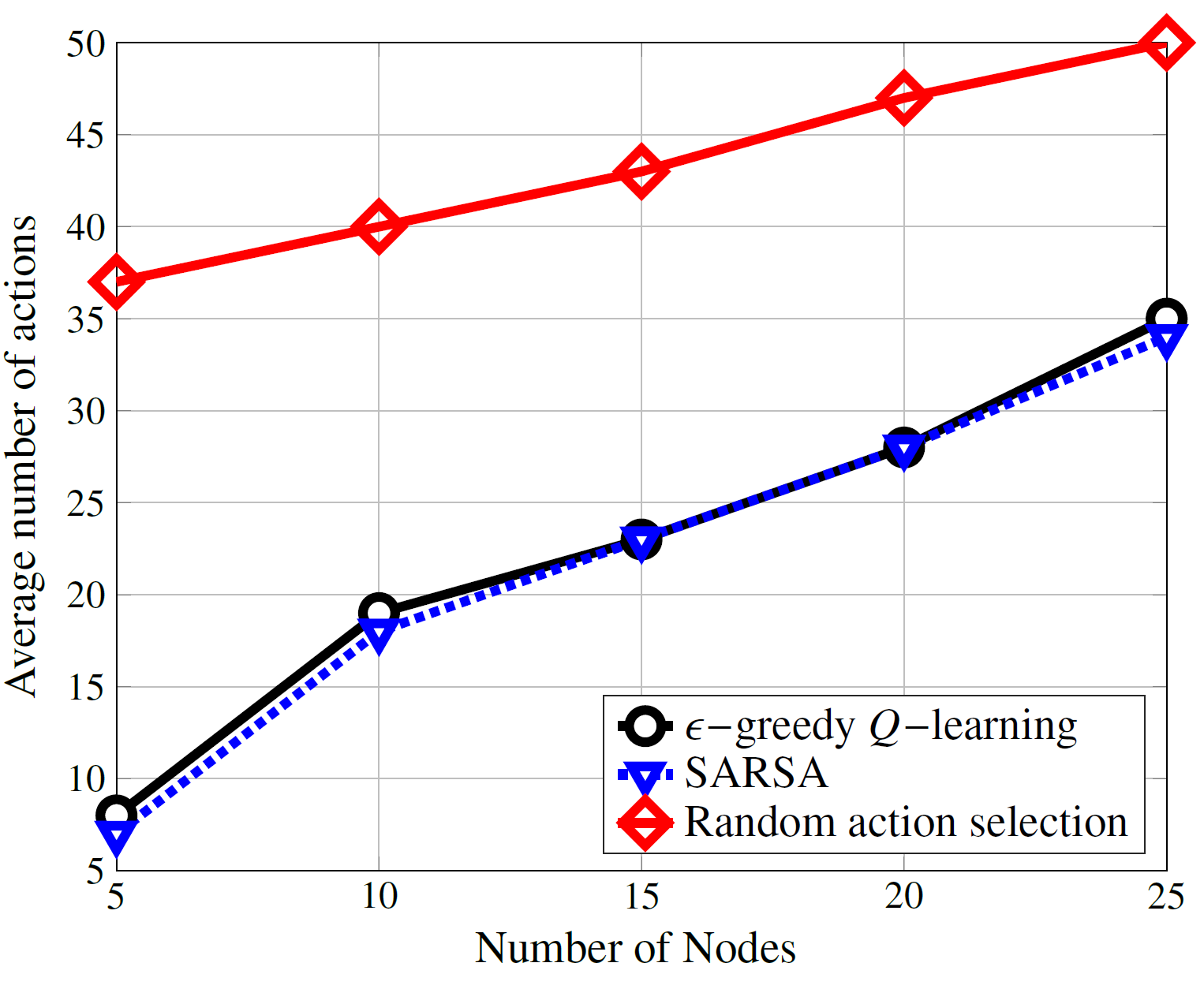}
    \caption{Average number of actions required to achieve a given magnitude of power harvesting. Clearly, randomly moving the AUV around the network requires far more steps to achieve a target throughput than $Q-$learning and SARSA. Moreover, the agent might never attain the maximum harvestable power when it takes random actions.}
    \label{fig:running_time_power}
\end{figure}

Figure~\ref{fig:harvest_power} shows the maximum harvested power for each algorithm at different network sizes.
When the PS factor $\alpha = 0$, all the received power is used for WPT.
% \hl{Similar story to the throughput analysis}
We note that while the trend for throughput is the same as that for WPT, the number of steps (actions) required for both shows marked differences, as more actions are required for WPT.
Figure~\ref{fig:running_time_power} shows how the number of actions changes for the different algorithms to achieve a target magnitude of power at the nodes.
% We observed that more actions are required for WPT than for throughput.
% Intuitively, this can be due to the fact that WPT requires more time to charge the supercapacitor bank of underwater nodes, unlike data transmission which requires only milliseconds.

\section{Conclusion}\label{conclusion}
This article presents an RL scheme designed to make underwater networks more sustainable.
The problem of jointly optimising throughput and harvested power was modelled as an MDP in a SWIPT-based underwater sensor network.
A $Q-$learning solution was developed to solve the MDP, based on SARSA and the $\epsilon-$greedy policies.
The proposed solution demonstrated high accuracy in finding optimal trajectories for maximising throughput and harvested power even in the rapidly changing underwater environment.
The solution was implemented in an AUV used for data collection and WPT, and tested in an open 3D RL environment developed for designed for this study. 
The environment is independent of the underlying channel model and can be used to evaluate any RL algorithms designed for underwater networks, irrespective of the signalling technology (acoustic, optical, magnetic induction or RF).
% Since this solution was the first known attempt at using RL for the intelligent implementation of SWIPT in underwater sensor networks, there was no benchmark scheme to compare the performance with.
% Instead, following a random trajectory was adopted as a simple baseline, whose performance was compared against  $Q-$learning and SARSA.
The performance results showed outstanding performance gaps between a random trajectory baseline model and the proposed scheme in terms of the ability to converge to a steady state, using throughput and energy efficiency as reward metrics.
The proposed scheme was also exploited to find the optimal PS ratio, $\alpha$, for allocating the received power for data transmissions and for WPT to the underwater sensor nodes.
%Our future work will involve a testbed implementation of the proposed scheme to evaluate its performance in real-world scenarios.

\bibliographystyle{IEEEtran}
\bibliography{main}

\begin{thebibliography}{10}
\providecommand{\url}[1]{#1}
\csname url@samestyle\endcsname
\providecommand{\newblock}{\relax}
\providecommand{\bibinfo}[2]{#2}
\providecommand{\BIBentrySTDinterwordspacing}{\spaceskip=0pt\relax}
\providecommand{\BIBentryALTinterwordstretchfactor}{4}
\providecommand{\BIBentryALTinterwordspacing}{\spaceskip=\fontdimen2\font plus
\BIBentryALTinterwordstretchfactor\fontdimen3\font minus
  \fontdimen4\font\relax}
\providecommand{\BIBforeignlanguage}[2]{{%
\expandafter\ifx\csname l@#1\endcsname\relax
\typeout{** WARNING: IEEEtran.bst: No hyphenation pattern has been}%
\typeout{** loaded for the language `#1'. Using the pattern for}%
\typeout{** the default language instead.}%
\else
\language=\csname l@#1\endcsname
\fi
#2}}
\providecommand{\BIBdecl}{\relax}
\BIBdecl

\bibitem{OverviewIoUTsDomingo2012}
M.~C. Domingo, ``An overview of the internet of underwater things,''
  \emph{Journal of Network and Computer Applications}, vol.~35, no.~6, pp.
  1879--1890, 2012.

\bibitem{IoUTsBigMarineDataJahanbakht2021}
M.~Jahanbakht, W.~Xiang, L.~Hanzo, and M.~Rahimi~Azghadi, ``Internet of
  underwater things and big marine data analytics—a comprehensive survey,''
  \emph{IEEE Communications Surveys \& Tutorials}, vol.~23, no.~2, pp.
  904--956, 2021.

\bibitem{RLIoUTs}
K.~G. Omeke, A.~I. Abubakar, L.~Zhang, Q.~H. Abbasi, and M.~A. Imran, ``How
  reinforcement learning is helping to solve internet-of-underwater-things
  problems,'' \emph{IEEE Internet of Things Magazine}, vol.~5, no.~4, pp.
  24--29, 2022.

\bibitem{SWIPTAnalysisBereketli2012}
A.~Bereketli and S.~Bilgen, ``Remotely powered underwater acoustic sensor
  networks,'' \emph{IEEE Sensors Journal}, vol.~12, no.~12, pp. 3467--3472,
  2012.

\bibitem{BatterylessIoUTGuida2022}
R.~Guida, E.~Demirors, N.~Dave, and T.~Melodia, ``Underwater ultrasonic
  wireless power transfer: A battery-less platform for the internet of
  underwater things,'' \emph{IEEE Transactions on Mobile Computing}, vol.~21,
  no.~5, pp. 1861--1873, 2022.

\bibitem{CapacityDistanceStojanovic2007}
M.~Stojanovic, ``On the relationship between capacity and distance in an
  underwater acoustic communication channel,'' \emph{ACM SIGMOBILE Mobile
  Computing and Communications Review}, vol.~11, no.~4, pp. 34--43, 2007.

\bibitem{AcousticContactlessChargingShahab2015}
S.~Shahab, M.~Gray, and A.~Erturk, ``An experimentally validated contactless
  acoustic energy transfer model with resistive-reactive electrical loading,''
  in \emph{Active and Passive Smart Structures and Integrated Systems 2015},
  vol. 9431.\hskip 1em plus 0.5em minus 0.4em\relax SPIE, 2015, pp. 31--44.

\bibitem{WirelessAcousticTransferKim2022}
H.~S. Kim, S.~Hur, D.-G. Lee, J.~Shin, H.~Qiao, S.~Mun, H.~Lee, W.~Moon,
  Y.~Kim, J.~M. Baik \emph{et~al.}, ``Ferroelectrically augmented contact
  electrification enables efficient acoustic energy transfer through liquid and
  solid media,'' \emph{Energy \& Environmental Science}, vol.~15, no.~3, pp.
  1243--1255, 2022.

\bibitem{PracticalAcousticWPTSi2013}
S.~H. SI, R.~Abd~Rahim \emph{et~al.}, ``Acoustic energy harvesting using
  piezoelectric generator for low frequency sound waves energy conversion,''
  \emph{International Journal of Engineering and Technology (IJET)}, 2013.

\bibitem{UltrasoundEnergyTransferShahab2014}
S.~Shahab, S.~Leadenham, F.~Guillot, K.~Sabra, and A.~Erturk, ``Ultrasound
  acoustic wave energy transfer and harvesting,'' in \emph{Active and Passive
  Smart Structures and Integrated Systems 2014}, vol. 9057.\hskip 1em plus
  0.5em minus 0.4em\relax SPIE, 2014, pp. 130--138.

\bibitem{RLTidalHarvestingHan2020}
M.~Han, J.~Duan, S.~Khairy, and L.~X. Cai, ``Enabling sustainable underwater
  iot networks with energy harvesting: A decentralized reinforcement learning
  approach,'' \emph{IEEE Internet of Things Journal}, vol.~7, no.~10, pp.
  9953--9964, 2020.

\bibitem{ShortestPathChargingAUVLin2018}
C.~Lin, K.~Wang, Z.~Chu, K.~Wang, J.~Deng, M.~S. Obaidat, and G.~Wu, ``Hybrid
  charging scheduling schemes for three-dimensional underwater wireless
  rechargeable sensor networks,'' \emph{Journal of Systems and Software}, vol.
  146, pp. 42--58, 2018.

\bibitem{UltrasonicTransducerDesignZhao2021}
Y.~Zhao, Y.~Du, Z.~Wang, J.~Wang, and Y.~Geng, ``Design of ultrasonic
  transducer structure for underwater wireless power transfer system,'' in
  \emph{2021 IEEE Wireless Power Transfer Conference (WPTC)}, 2021, pp. 1--4.

\bibitem{AcousticSWIPTNOMAEsmaiel2020}
H.~Esmaiel, Z.~A. Qasem, H.~Sun, J.~Qi, J.~Wang, and Y.~Gu, ``Wireless
  information and power transfer for underwater acoustic time-reversed noma,''
  \emph{IET Communications}, vol.~14, no.~19, pp. 3394--3403, 2020.

\bibitem{WPTAUVsWang2023}
Y.~Wang, T.~Li, M.~Zeng, J.~Mai, P.~Gu, and D.~Xu, ``An underwater simultaneous
  wireless power and data transfer system for auv with high-rate full-duplex
  communication,'' \emph{IEEE Transactions on Power Electronics}, vol.~38,
  no.~1, pp. 619--633, 2023.

\bibitem{ReviewWirelessAUVChargingTeeneti2021}
C.~R. Teeneti, T.~T. Truscott, D.~N. Beal, and Z.~Pantic, ``Review of wireless
  charging systems for autonomous underwater vehicles,'' \emph{IEEE Journal of
  Oceanic Engineering}, vol.~46, no.~1, pp. 68--87, 2021.

\bibitem{ConductiveCouplerUnderwaterTamura2021}
M.~Tamura, K.~Murai, and M.~Matsumoto, ``Design of conductive coupler for
  underwater wireless power and data transfer,'' \emph{IEEE Transactions on
  Microwave Theory and Techniques}, vol.~69, no.~1, pp. 1161--1175, 2021.

\bibitem{MIMultiAUVChargingGuo2021}
H.~Guo, Z.~Sun, and P.~Wang, ``Joint design of communication, wireless energy
  transfer, and control for swarm autonomous underwater vehicles,'' \emph{IEEE
  Transactions on Vehicular Technology}, vol.~70, no.~2, pp. 1821--1835, 2021.

\bibitem{InductiveCouplingWPTLin2019}
R.~Lin, D.~Li, T.~Zhang, and M.~Lin, ``A non-contact docking system for
  charging and recovering autonomous underwater vehicle,'' \emph{Journal of
  Marine Science and Technology}, vol.~24, no.~3, pp. 902--916, 2019.

\bibitem{SEANetDemirors2016}
E.~Demirors, J.~Shi, R.~Guida, and T.~Melodia, ``Seanet g2: Toward a
  high-data-rate software-defined underwater acoustic networking platform,'' in
  \emph{Proceedings of the 11th ACM International Conference on Underwater
  Networks \& Systems}, 2016, pp. 1--8.

\bibitem{InductivePowerTransferCheng2014}
Z.~Cheng, Y.~Lei, K.~Song, and C.~Zhu, ``Design and loss analysis of loosely
  coupled transformer for an underwater high-power inductive power transfer
  system,'' \emph{IEEE Transactions on Magnetics}, vol.~51, no.~7, pp. 1--10,
  2014.

\bibitem{SoundPrinciplesUrick1975}
R.~J. Urick, \emph{Principles of underwater sound-2}.\hskip 1em plus 0.5em
  minus 0.4em\relax New York, NY (USA) McGraw-Hill Book, 1975.

\bibitem{ThorpFormula1967}
W.~H. Thorp, ``Analytic description of the low-frequency attenuation
  coefficient,'' \emph{The Journal of the Acoustical Society of America},
  vol.~42, no.~1, pp. 270--270, 1967.

\bibitem{AppliedUnderwaterAcousticsBjorno2017}
L.~Bj{\o}rn{\o}, T.~Neighbors, and D.~Bradley, \emph{Applied underwater
  acoustics}.\hskip 1em plus 0.5em minus 0.4em\relax Elsevier, 2017.

\bibitem{AUVEnergy2020}
K.~G. Omeke, M.~S. Mollel, L.~Zhang, Q.~H. Abbasi, and M.~A. Imran, ``Energy
  optimisation through path selection for underwater wireless sensor
  networks,'' in \emph{2020 International Conference on UK-China Emerging
  Technologies (UCET)}, 2020, pp. 1--4.

\bibitem{AUVGuidance}
\BIBentryALTinterwordspacing
F.~Scibilia, U.~Jørgensen, and R.~Skjetne, ``Auv guidance system for dynamic
  trajectory generation,'' \emph{IFAC Proceedings Volumes}, vol.~45, no.~5, pp.
  198 -- 203, 2012, 3rd IFAC Workshop on Navigation, Guidance and Control of
  Underwater Vehicles. [Online]. Available:
  \url{http://www.sciencedirect.com/science/article/pii/S1474667016306024}
\BIBentrySTDinterwordspacing

\bibitem{AUVPlatforms}
\BIBentryALTinterwordspacing
J.~Bellingham, ``Platforms: Autonomous underwater vehicles,'' in
  \emph{Encyclopedia of Ocean Sciences (Second Edition)}, J.~H. Steele,
  Ed.\hskip 1em plus 0.5em minus 0.4em\relax Oxford: Academic Press, 2009, pp.
  473 -- 484. [Online]. Available:
  \url{http://www.sciencedirect.com/science/article/pii/B978012374473900730X}
\BIBentrySTDinterwordspacing

\bibitem{OfflineRLSurveyPrudencio2022}
R.~F. Prudencio, M.~R. Maximo, and E.~L. Colombini, ``A survey on offline
  reinforcement learning: Taxonomy, review, and open problems,'' \emph{arXiv
  preprint arXiv:2203.01387}, 2022.

\bibitem{RLBookSutton2018}
R.~S. Sutton and A.~G. Barto, \emph{Reinforcement learning: An
  introduction}.\hskip 1em plus 0.5em minus 0.4em\relax MIT press, 2018.

\bibitem{PolicyBasedRLSewak2019}
M.~Sewak, ``Policy-based reinforcement learning approaches,'' in \emph{Deep
  Reinforcement Learning}.\hskip 1em plus 0.5em minus 0.4em\relax Springer,
  2019, pp. 127--140.

\bibitem{StatisticalChannelModellingQarabaqi2013}
P.~Qarabaqi and M.~Stojanovic, ``Statistical characterization and
  computationally efficient modeling of a class of underwater acoustic
  communication channels,'' \emph{IEEE Journal of Oceanic Engineering},
  vol.~38, no.~4, pp. 701--717, 2013.

\bibitem{UWAChannelModelingMorozs2020}
N.~Morozs, W.~Gorma, B.~T. Henson, L.~Shen, P.~D. Mitchell, and Y.~V. Zakharov,
  ``Channel modeling for underwater acoustic network simulation,'' \emph{IEEE
  Access}, vol.~8, pp. 136\,151--136\,175, 2020.

\bibitem{MDPsEHFadingChannelsLi2015}
W.~Li, M.-L. Ku, Y.~Chen, and K.~J.~R. Liu, ``On outage probability for
  stochastic energy harvesting communications in fading channels,'' \emph{IEEE
  Signal Processing Letters}, vol.~22, no.~11, pp. 1893--1897, 2015.

\end{thebibliography}
\end{document}